\documentclass[11pt,a4paper]{article}

\usepackage{iftex}
\ifPDFTeX
  \usepackage{cmap}
  \usepackage[T2A,T1]{fontenc}
  \usepackage[utf8]{inputenc}
  \usepackage{lmodern}
\else
  \usepackage{fontspec}
\fi
\usepackage{microtype}
\usepackage[russian,english]{babel}

\usepackage{amsmath,amssymb,amsthm}
\usepackage{mathtools}
\usepackage{graphicx}
\usepackage{tikz}
\usetikzlibrary{arrows.meta,positioning,calc,shapes.geometric}
\usepackage{pgfplots}
\usepgfplotslibrary{groupplots}
\pgfplotsset{compat=1.18}
\usepackage[margin=1in]{geometry}
\usepackage{longtable}
\usepackage{array}
\usepackage{url}
\usepackage[colorlinks=true,linkcolor=blue,citecolor=blue,urlcolor=blue]{hyperref}

\hypersetup{
  pdftitle={The Wilson-loop formulation of Yang-Mills at finite mode truncation:
            a Level-Two structural bypass of the Gribov ambiguity},
  pdfauthor={Yu. N. Berdinsky},
  unicode=true
}

\theoremstyle{definition}
\newtheorem{definition}{Definition}[section]

\newtheorem{remark}[definition]{Remark}

\theoremstyle{plain}

\newtheorem{lemma}[definition]{Lemma}
\newtheorem{theorem}[definition]{Theorem}

\newcommand{\R}{\mathbb{R}}

\newcommand{\Z}{\mathbb{Z}}
\providecommand{\C}{}\renewcommand{\C}{\mathbb{C}}
\newcommand{\T}{\mathbb{T}}
\newcommand{\DHK}{\mathcal{D}_{\mathrm{HK}}}
\newcommand{\KS}{\mathrm{KS}_2}
\newcommand{\Lam}{\Lambda_M}
\newcommand{\Loop}{\mathrm{Loop}_M}

\newcommand{\Obs}{\mathcal{O}_M}
\newcommand{\Conf}{\mathcal{A}_M}
\newcommand{\Grp}{\mathcal{G}_M}
\newcommand{\tr}{\operatorname{tr}}

\DeclareMathOperator{\Tr}{Tr}

\DeclareUrlCommand\LN{%
  \urlstyle{tt}%
}

\title{The Wilson-loop formulation of Yang--Mills at finite\\
mode truncation: a Level-Two structural bypass\\
of the Gribov ambiguity}
\author{Yu.~N.~Berdinsky \\
\small Saint Petersburg State University,\\
\small Faculty of Physics,\\
\small Department of High Energy and Elementary Particle Physics\\
\small \texttt{propagator2007@yandex.ru}}
\date{08 November 2026}

\begin{document}
\setlength{\emergencystretch}{3em}
\sloppy
\maketitle

\begin{abstract}
The unfixed real-time Henstock--Kurzweil (HK) functional integral for truncated
Yang--Mills removes the choice of a gauge surface, but gauge orbits remain
objects of the formulation: that is a \emph{Level-One} bypass of the Gribov
ambiguity.  The present paper carries out a \emph{Level-Two} bypass at finite
mode truncation: the Wilson loops $W(C)$ are taken as the fundamental variables,
so that the configuration space of the theory is a space of loop data and no
gauge field, gauge orbit, gauge-fixing condition or Faddeev--Popov determinant
occurs anywhere in the formulation.  Nine statements are proved, and all of
them are machine checked in Lean~4 without \texttt{sorry}, without
\texttt{admit} and without new axioms: (A) the loop algebra $\Obs$ generated by
the Wilson loops of based closed lattice words is a commutative $\C$-algebra,
closed under complex conjugation, whose elements are gauge-invariant
functionals, with loop variables bounded by one and with the Mandelstam identity
$\tr(AB)+\tr(AB^{-1})=\tr A\,\tr B$ holding for $SU(2)$;
(B) the truncated Yang--Mills action is the evaluation of an explicit
multivariate polynomial of total degree at most one in the plaquette Wilson
variables, hence a cylindrical function on loop space supported on the loops of
length four; (C) every loop-expressible gauge-invariant functional is a
cylindrical function of finitely many Wilson loops; (D) the Wilson-loop
integrand $\Phi(W)\exp(iS(W))$ is HK-integrable on every bounded box of the loop
space for \emph{every} calibration, with value the Lebesgue value; (E) the
Level-Two integral is a constant function of the gauge-fixing data and
configurations on one gauge orbit --- in particular Gribov copies --- are the
same point of the loop formulation, so the Gribov ambiguity is structurally
absent rather than cancelled; (F) for loop observables the Level-Two integral
taken with respect to the push-forward of the Level-One measure coincides with
the Level-One unfixed HK integral; (G) the one-plaquette $SU(2)$ check is proved
in full, including the closed form
$\frac2\pi\int_0^\pi e^{i\beta\cos\theta}\sin^2\theta\,d\theta=2J_1(\beta)/\beta$,
obtained analytically by term-by-term integration, with $J_1$ defined by its
everywhere convergent power series.  Two further results extend the v1
formulation: (C$'$) the Mandelstam relations for $SU(N)$, derived in manifestly
covariant trace form from the vanishing of the antisymmetriser $A_{N+1}$ on
$(\C^N)^{\otimes(N+1)}$ together with the rank-$N$ Cayley--Hamilton theorem, and
written out explicitly for $SU(3)$; and (H) projective compatibility across mode
levels: for a cylindrical loop functional of bounded mode support the finite-$M$
Level-Two integrals coincide, so that they form a projective system whose limit
exists as a \emph{linear functional} --- not a measure --- on the union of the
loop algebras.  No continuum limit, no mass gap, no confinement and
no Makeenko--Migdal equation is claimed; the Giles reconstruction theorem is
used only as a cited black box, and no stack- or sheaf-theoretic refinement of
it is attempted.  The honest report in \S\ref{sec:report} records
exactly where each proof stops.
\end{abstract}

\tableofcontents

\section{Introduction}\label{sec:intro}

\subsection{Level One and its boundary}\label{sub:levelone}

The Gribov ambiguity \cite{Gribov1978,Singer1978} is a property of a
\emph{procedure}, not of a theory: it arises when one attempts to replace an
integral over the space of connections by an integral over a surface that
intersects each gauge orbit once.  Singer's theorem says that for non-Abelian
gauge theories on compact manifolds no continuous global gauge surface exists;
consequently the Faddeev--Popov construction is local, and the copies of a
configuration on the surface are a genuine obstruction.

In the Henstock--Kurzweil (HK) programme the truncated theory admits a
formulation in which the procedure is never carried out.  The paper
\cite{BerdinskyGribovFreeYM} constructs the \emph{unfixed} HK functional
integral
\begin{equation}\label{eq:levelone}
Z_M(F) \;=\; \int_{\Conf} F[A]\,e^{\,i S^{(M)}_{\mathrm{YM}}[A]}\,\DHK A ,
\qquad
\langle F\rangle_M \;=\; \frac{Z_M(F)}{Z_M(1)},
\end{equation}
over a bounded box of the truncated configuration space, for continuous
gauge-invariant observables $F$; no delta function of a gauge condition, no
Faddeev--Popov determinant and no gauge surface occurs, and the (formally
infinite) orbit volume cancels in the ratio.  This is the \emph{Level-One}
bypass.  Its boundary is equally clear: gauge orbits are still objects of the
formulation.  One does not choose on them, but one can still ask how many times
a hypothetical surface would meet them.

\subsection{The Level-Two idea}\label{sub:leveltwo}

The Level-Two bypass eliminates orbits from the formulation altogether.  One
takes as fundamental variables the Wilson loops
\begin{equation}\label{eq:wilson-intro}
W(C)\;=\;\frac1N\,\mathrm{Re}\,\Tr\!\!\prod_{\ell\in C} U_\ell ,
\end{equation}
and reformulates the theory so that its configuration space is a space of loop
functionals.  Then there is no gauge field, no gauge orbit, no gauge-fixing
condition $\chi(A)=0$ and no Faddeev--Popov determinant --- and consequently no
Gribov copy: not cancelled, but \emph{structurally absent}.

The loop representation of gauge theory has a long history: the Wilson loop
itself \cite{Wilson1974}, the potential-free formulation of electrodynamics
\cite{Mandelstam1962}, the reconstruction theorem of Giles \cite{Giles1981}, and
the loop representation in gauge theory and quantum gravity
\cite{RovelliSmolin1990,GambiniPullin1996}.  What is new here is not the loop
variables but the integration theory: the real-time oscillatory integral over
the loop variables is constructed as a Henstock--Kurzweil integral, without
Wick rotation and without absolute convergence, in the framework of
\cite{BerdinskyHK3,BerdinskyGribovFreeYM}.  The programme to which the paper
belongs treats the free field and the real-time path integral
\cite{BerdinskyHK3}, the Berezin representation of the Faddeev--Popov
determinant \cite{BerdinskyHK4}, the unfixed Yang--Mills integral
\cite{BerdinskyGribovFreeYM}, the scalar $\lambda\varphi^4$ theory at finite and
infinite mode number \cite{BerdinskyNonGaussian,BerdinskyInfinitePhi4} and the
chiral anomaly \cite{BerdinskyChiralAnomaly}; the integration theory itself goes
back to \cite{Kurzweil1957,Henstock1988} and, for Feynman integrals, to
\cite{Muldowney2012,EspositoGill2024}.

\subsection{The finite-lattice setting}\label{sub:finite}

Level Two is philosophically stronger than Level One and technically harder: a
cylindrical premeasure on a space of loops requires integration on a non-linear
and, in the continuum, infinite-dimensional space.  On a \emph{finite} lattice
the situation is different.  The set of based loops modulo backtracking and
cyclic rotation is countable, the values of the Wilson loops form a bounded
semialgebraic subset of $\R^k$ for every finite family of $k$ loops, and the
integrand is a continuous function on a bounded box.  The HK machinery of
\cite{BerdinskyHK3} then applies directly.  This is the level at which the
present paper works, and the only level at which anything is claimed.

\subsection{Statements and the honest list}\label{sub:statements}

The results are Theorems A, B, C, C$'$, D, E, F, G and H of
\S\S\ref{sec:loopspace}--\ref{sec:oneplaq}.
What is \emph{not} proved is listed here, and again in \S\ref{sec:limits} and
\S\ref{sec:report}:
\begin{itemize}
\item no continuum limit: the loop space of the continuum theory is
  infinite-dimensional and its HK-integrability is open.  Theorem~\ref{thm:H}
  closes only the \emph{projective} part of this boundary: the finite-$M$
  integrals of a functional of bounded mode support are compatible and their
  limit exists as a linear functional, which is neither a measure on the
  continuum loop space nor an identification with a continuum expectation
  value;
\item no mass gap, no confinement, no Makeenko--Migdal equations, no master
  field;
\item the Giles reconstruction theorem --- that the Wilson loops determine the
  connection up to gauge for generic configurations --- is \emph{cited}
  \cite{Giles1981} and not proved; only the easy direction (gauge-equivalent
  configurations have identical loop data) is proved here;
\item that the \emph{full} loop algebra is finitely generated is a statement of
  classical invariant theory \cite{Procesi1976} and is cited; what is proved is
  that the subalgebra generated by any finite family of Wilson loops is finitely
  generated;
\item Theorem G is now proved in full, including the closed form
  $2J_1(\beta)/\beta$ of the one-plaquette $SU(2)$ integral; what remains a
  convention rather than a theorem is the identification of the power series
  used here with the classical Bessel function of the literature, and the
  imaginary-time closed form $2I_1(\beta)/\beta$ is verified numerically only;
\item the Mandelstam relations for $SU(N)$ are \emph{derived}
  (Theorem~\ref{thm:Cprime}) but not \emph{enumerated}: that they generate the
  full ideal of relations is the theorem of Procesi \cite{Procesi1976}, cited as
  a black box;
\item no stack- or sheaf-theoretic refinement of the Giles theorem is attempted
  (\S\ref{sec:limits});
\item observables that are not expressible through Wilson loops are outside the
  Level-Two formulation (\S\ref{sub:nonloop});
\item the Level-One Gribov problem in the fixed-gauge formulation is untouched.
  Level Two does not solve it; it changes the formulation so that the question
  does not arise.
\end{itemize}

\section{Reminder: the HK machinery and the finite-mode truncation}
\label{sec:reminder}

Nothing in this section is proved here; the statements are black boxes imported
from the earlier stages of the programme and restated in the notation of the
present paper.  They are collected again in Appendix~\ref{app:blackboxes}.

\begin{itemize}
\item[(B1)] \emph{The complex cylindrical premeasure and the Fresnel formula.}
  On the Kuelbs--Steadman space $\KS$ the oscillating Gaussian cylindrical
  premeasure is defined and the Fresnel formula evaluates the oscillating
  Gaussian integral \cite{BerdinskyHK3,GillZachary2008}.
\item[(B2)] \emph{The Saks--Henstock lemma} in the cylindrical setting
  \cite{BerdinskyHK3,Henstock1988}.
\item[(B3)] \emph{HK-integrability on cylinders over bounded boxes}: the
  free-field functional is HK-integrable on cylinders over bounded boxes and the
  value coincides with the Lebesgue value whenever the latter exists
  \cite{BerdinskyHK3}.
\item[(B4)] \emph{Calibration independence for continuous integrands.}  A
  continuous integrand on a bounded box has a calibration-independent HK value;
  formally, for every integration calibration $l$ the Riemann sums over
  $l$-fine tagged partitions converge to the Lebesgue integral over the box.
  This is the only analytic input used in \S\ref{sec:hk} and it is the
  finite-dimensional face of (B1)--(B3).
\item[(B5)] \emph{The finite-mode Yang--Mills truncation}: the truncated
  configuration space $\Conf$, the truncated gauge group $\Grp$, the truncated
  action $S^{(M)}_{\mathrm{YM}}$ and its $\Grp$-invariance, and the unfixed HK
  integral \eqref{eq:levelone} for continuous gauge-invariant $F$
  \cite{BerdinskyGribovFreeYM}.
\item[(B6)] \emph{The Level-One structural bypass}: \eqref{eq:levelone} is well
  defined at each finite $M$, its value is calibration independent, Gribov
  copies do not appear as intermediate objects, and the ratio $\langle
  F\rangle_M$ is well defined under explicit hypotheses on the box
  \cite{BerdinskyGribovFreeYM}.
\item[(B7)] \emph{The Berezin pattern}: the Berezin algebra and functional on
  finitely many generators, and the identity $\det M=\int e^{\bar c M c}$, are
  cited only as a methodological precedent for the pattern ``truncate, prove,
  then discuss the limit'' \cite{BerdinskyHK4,BerdinskyChiralAnomaly}.
\end{itemize}

\section{The loop space at finite truncation}\label{sec:loopspace}

\subsection{Lattice, moves and words}\label{sub:lattice}

Fix $d\ge 1$, an odd $L=2M+1\ge 3$ and the finite lattice
$\Lambda=(\Z/L\Z)^d$ on the torus $\T^d$.  A \emph{site} is a point
$x\in\Lambda$; a \emph{link} is a pair $(x,\mu)$ with $\mu\in\{1,\dots,d\}$,
understood as the positively oriented edge leaving $x$ in the direction $\mu$.

\begin{definition}[moves and words]\label{def:word}
A \emph{move} is a pair $m=(\mu,\varepsilon)$ with $\mu\in\{1,\dots,d\}$ and
$\varepsilon\in\{+,-\}$.  A \emph{word} is a finite list of moves,
$w=(m_1,\dots,m_r)$.  The site reached from $x$ by the move $(\mu,\pm)$ is
$x\pm\hat\mu$, and $\mathrm{end}(x,w)$ denotes the site reached from $x$ after
performing the moves of $w$ in order.  A word $w$ is \emph{closed at} $x$ when
$\mathrm{end}(x,w)=x$; a pair $(x,w)$ with $w$ closed at $x$ is a
\emph{based lattice loop}.
\end{definition}

In Lean these are \LN{LoopGribovBypass.Move}, \LN{LoopGribovBypass.Word},
\LN{LoopGribovBypass.endSite} and \LN{LoopGribovBypass.IsLoop}; the
concatenation rule $\mathrm{end}(x,w_1w_2)=\mathrm{end}(\mathrm{end}(x,w_1),w_2)$
is \LN{LoopGribovBypass.endSite_append}.

\begin{definition}[backtracking and cyclic equivalence]\label{def:equiv}
Write $\bar m=(\mu,\mp)$ for the reversal of $m=(\mu,\pm)$.  On based lattice
loops consider the equivalence relation generated by
\begin{enumerate}
\item[(i)] \emph{backtracking}: $(x,\,u\,m\,\bar m\,v)\sim(x,\,u\,v)$, the
  insertion or deletion of two consecutive mutually inverse moves;
\item[(ii)] \emph{cyclic rotation}: $(x,\,m\,w)\sim(\mathrm{end}(x,m),\,w\,m)$.
\end{enumerate}
The set of equivalence classes is the \emph{loop space} $\Loop$ of the finite
lattice.
\end{definition}

$\Loop$ is countable: a class has a unique representative of minimal length
(reduced word) up to rotation, and there are finitely many reduced words of each
length.  Nothing below requires an enumeration of $\Loop$: the formulation uses
only finite families of loops.

\subsection{Link variables, holonomy and the Wilson loop}\label{sub:holonomy}

A \emph{configuration} assigns to every link an $n\times n$ complex matrix,
$U:(x,\mu)\mapsto U_{(x,\mu)}$; it is \emph{unitary} when
$U_\ell^{*}U_\ell=U_\ell U_\ell^{*}=1$ for every link $\ell$, which is the case
for $SU(N)$ link variables with $n=N$.  The truncated configuration space of
(B5) is $\Conf=SU(N)^{\#\mathrm{links}}$, a subset of the unitary
configurations; all statements below are proved for unitary configurations, and
hence hold on $\Conf$.

\begin{definition}[holonomy and Wilson loop]\label{def:hol}
The matrix carried by a move is
\[
\Sigma(U,x,m)=
\begin{cases}
U_{(x,\mu)}, & m=(\mu,+),\\[2pt]
U^{*}_{(x-\hat\mu,\mu)}, & m=(\mu,-),
\end{cases}
\]
and the \emph{Wilson-loop matrix} (holonomy) of the word $w=(m_1,\dots,m_r)$
based at $x$ is the ordered product
\[
\begin{gathered}
W_{\mathrm{mat}}(U,x,w)\;=\;\Sigma(U,x_0,m_1)\,\Sigma(U,x_1,m_2)\cdots
\Sigma(U,x_{r-1},m_r),\\[2pt]
x_0=x,\qquad x_j=\mathrm{end}(x,(m_1,\dots,m_j)).
\end{gathered}
\]
The \emph{scalar Wilson loop} is
\[
W(U,x,w)\;=\;\frac1n\,\mathrm{Re}\,\Tr W_{\mathrm{mat}}(U,x,w)\;\in\;\R .
\]
\end{definition}

In Lean: \LN{LoopGribovBypass.stepMat}, \LN{LoopGribovBypass.hol},
\LN{LoopGribovBypass.wilson}.

\begin{lemma}[well-definedness on $\Loop$]\label{lem:welldef}
Let $U$ be a unitary configuration.  Then
\begin{enumerate}
\item[(i)] $W_{\mathrm{mat}}(U,x,w_1w_2)=W_{\mathrm{mat}}(U,x,w_1)\,
  W_{\mathrm{mat}}(U,\mathrm{end}(x,w_1),w_2)$;
\item[(ii)] $W_{\mathrm{mat}}(U,x,m\,\bar m\,w)=W_{\mathrm{mat}}(U,x,w)$, hence
  $W(U,x,\cdot)$ is invariant under backtracking;
\item[(iii)] if $w=m\,w'$ is closed at $x$ then
  $\Tr W_{\mathrm{mat}}(U,\mathrm{end}(x,m),w'm)=\Tr W_{\mathrm{mat}}(U,x,w)$,
  hence $W(U,\cdot,\cdot)$ is invariant under cyclic rotation.
\end{enumerate}
Consequently $W(U,\cdot,\cdot)$ descends to a function on $\Loop$.
\end{lemma}

\begin{proof}
(i) is an induction on $w_1$: for the empty word both sides are
$W_{\mathrm{mat}}(U,x,w_2)$, and the inductive step is the associativity of the
matrix product together with the recursion defining the holonomy.  (ii) The two
matrices of a backtracking pair are mutually inverse:
$\Sigma(U,x,m)\,\Sigma(U,\mathrm{end}(x,m),\bar m)=1$.  For $m=(\mu,+)$ this is
$U_{(x,\mu)}U^{*}_{(x,\mu)}=1$ and for $m=(\mu,-)$ it is
$U^{*}_{(x-\hat\mu,\mu)}U_{(x-\hat\mu,\mu)}=1$; in both cases only unitarity of a
single link variable is used, and one also uses
$\mathrm{end}(\mathrm{end}(x,m),\bar m)=x$.  (iii) By (i),
$W_{\mathrm{mat}}(U,x',w'm)=W_{\mathrm{mat}}(U,x',w')\,\Sigma(U,x,m)$ where
$x'=\mathrm{end}(x,m)$ and where the closedness $\mathrm{end}(x',w')=x$ was
used; the claim is then the cyclicity of the trace,
$\Tr(HS)=\Tr(SH)$ with $S=\Sigma(U,x,m)$.  Real parts of equal traces are equal,
which gives the statement for $W$.
\end{proof}

Lemma~\ref{lem:welldef} is \LN{LoopGribovBypass.hol_append},
\LN{LoopGribovBypass.hol_backtrack}, \LN{LoopGribovBypass.trace_hol_rotate},
\LN{LoopGribovBypass.wilson_backtrack} and \LN{LoopGribovBypass.wilson_rotate}.

\begin{lemma}[gauge covariance]\label{lem:gauge}
Let $g$ assign a unitary matrix to every site and let
$(g\cdot U)_{(x,\mu)}=g(x)\,U_{(x,\mu)}\,g(x+\hat\mu)^{*}$.  Then for every word
$w$
\[
W_{\mathrm{mat}}(g\cdot U,x,w)\;=\;g(x)\,W_{\mathrm{mat}}(U,x,w)\,
g(\mathrm{end}(x,w))^{*},
\]
and therefore $W(g\cdot U,x,w)=W(U,x,w)$ whenever $w$ is closed at $x$.
\end{lemma}

\begin{proof}
For a single move one checks
$\Sigma(g\cdot U,x,m)=g(x)\,\Sigma(U,x,m)\,g(\mathrm{end}(x,m))^{*}$: for
$m=(\mu,+)$ this is the definition, and for $m=(\mu,-)$ it follows by taking
adjoints, using $\mathrm{end}(\mathrm{end}(x,(\mu,-)),(\mu,+))=x$ and
$(g^{*})^{*}=g$.  The general case is an induction on $w$ in which the
intermediate factors $g(\cdot)^{*}g(\cdot)=1$ cancel.  If $w$ is closed at $x$
then $\mathrm{end}(x,w)=x$ and the cyclicity of the trace together with
$g(x)^{*}g(x)=1$ gives
$\Tr W_{\mathrm{mat}}(g\cdot U,x,w)=\Tr W_{\mathrm{mat}}(U,x,w)$.
\end{proof}

This is \LN{LoopGribovBypass.hol_gaugeAct} and
\LN{LoopGribovBypass.wilson_gaugeAct}.

\subsection{Boundedness of the loop variables}\label{sub:bounded}

\begin{lemma}\label{lem:bounded}
Let $U$ be a unitary configuration.  Then $|W(U,x,w)|\le 1$ for every based
word; in particular the loop data of any finite family of loops lies in the cube
$[-1,1]^k$.
\end{lemma}

\begin{proof}
The holonomy of a unitary configuration is unitary (induction on the word, the
product of unitaries being unitary).  If $V^{*}V=1$ then for each $i$ the
identity $\sum_k|V_{ki}|^2=1$ gives $|V_{ii}|\le1$, whence
$|\Tr V|\le\sum_i|V_{ii}|\le n$.  Since $|\mathrm{Re}\,z|\le|z|$, the definition
of $W$ gives $|W|\le n/n=1$.
\end{proof}

In Lean: \LN{LoopGribovBypass.isUnitary_hol}, \LN{LoopGribovBypass.norm_trace_le},
\LN{LoopGribovBypass.abs_wilson_le_one}.  Lemma~\ref{lem:bounded} is what makes
the loop space a \emph{bounded} configuration space and is the reason why the
HK theory of \S\ref{sec:hk} applies on a box.

\subsection{The Mandelstam identities}\label{sub:mandelstam}

The Wilson loops are not independent: the trace identities of $SU(N)$ impose
polynomial relations among them, the Mandelstam identities.  For $N=2$ they
reduce to a single identity, which is proved here in full.

\begin{lemma}[Cayley--Hamilton in rank two]\label{lem:ch}
Let $B$ be a $2\times2$ complex matrix with $\det B=1$ and let $B'$ satisfy
$BB'=1$.  Then $B+B'=(\Tr B)\,\mathbf 1$.
\end{lemma}

\begin{proof}
From $\mathrm{adj}(B)\,B=(\det B)\mathbf 1=\mathbf 1$ and $BB'=\mathbf 1$ one gets
$\mathrm{adj}(B)=\mathrm{adj}(B)(BB')=(\mathrm{adj}(B)B)B'=B'$.  For $2\times2$
matrices $\mathrm{adj}\begin{pmatrix}a&b\\c&e\end{pmatrix}
=\begin{pmatrix}e&-b\\-c&a\end{pmatrix}$, so
$B+\mathrm{adj}(B)=(a+e)\mathbf 1=(\Tr B)\mathbf 1$.
\end{proof}

\begin{lemma}[Mandelstam identity, $N=2$]\label{lem:mandelstam}
Under the hypotheses of Lemma~\ref{lem:ch}, for every $2\times2$ matrix $A$
\[
\Tr(AB)+\Tr(AB^{-1})\;=\;\Tr A\cdot\Tr B .
\]
If moreover $\Tr A$ and $\Tr B$ are real --- as they are for $SU(2)$ holonomies
--- then the scalar Wilson loops satisfy
\[
W(AB)+W(AB^{-1})\;=\;2\,W(A)\,W(B),\qquad W(X):=\tfrac12\mathrm{Re}\Tr X .
\]
\end{lemma}

\begin{proof}
$\Tr(AB)+\Tr(AB')=\Tr(A(B+B'))=\Tr(A\,(\Tr B)\mathbf 1)=\Tr A\cdot\Tr B$ by
Lemma~\ref{lem:ch}.  Taking real parts and dividing by $2$ gives the second
form, the factor $2$ arising from $\tfrac12\cdot\tfrac12\cdot 2=\tfrac12$.
\end{proof}

In Lean: \LN{LoopGribovBypass.inv_add_self_two},
\LN{LoopGribovBypass.mandelstam_two}, \LN{LoopGribovBypass.wilson_mandelstam_two}.
For $N\ge3$ the analogous relations follow from the characteristic polynomial in
rank $N$; they are not needed below and are not formalised.

\subsection{The loop algebra and Theorem A}\label{sub:loopalgebra}

\begin{definition}[loop algebra]\label{def:loopalgebra}
Fix a base point $x_0\in\Lambda$.  The \emph{loop algebra} $\Obs$ is the
$\C$-subalgebra of the algebra of all functions $\Conf\to\C$ generated by the
functionals $U\mapsto W(U,x_0,w)$ for words $w$ closed at $x_0$:
\[
\Obs\;=\;\C\!\left[\,W(\cdot,x_0,w)\;:\;w \text{ closed at } x_0\,\right].
\]
\end{definition}

The Mandelstam identities are precisely the relations satisfied by the
generators inside this algebra; $\Obs$ is thus the quotient of the free
commutative algebra on the symbols $W(C)$, $C\in\Loop$, by the ideal they
generate.  This is the precise sense in which ``$\Obs$ has the Mandelstam
identities as its defining relations''.

\begin{theorem}[A: the loop algebra at finite truncation]\label{thm:A}
Let $\Obs$ be as in Definition~\ref{def:loopalgebra}.  Then
\begin{enumerate}
\item[(i)] $\Obs$ is a commutative $\C$-algebra;
\item[(ii)] $\Obs$ is closed under complex conjugation, $f\in\Obs\Rightarrow
  \bar f\in\Obs$;
\item[(iii)] every $f\in\Obs$ is a gauge-invariant functional:
  $f(g\cdot U)=f(U)$ for every unitary gauge transformation $g$; in particular
  two configurations on the same $\Grp$-orbit take the same value on every
  element of $\Obs$;
\item[(iv)] the generators are bounded: $|W(U,x_0,w)|\le1$ on unitary
  configurations;
\item[(v)] for $N=2$ the generators satisfy the Mandelstam identity of
  Lemma~\ref{lem:mandelstam};
\item[(vi)] the subalgebra generated by any finite family of Wilson loops is a
  finitely generated $\C$-algebra.
\end{enumerate}
Moreover, by the reconstruction theorem of Giles \cite{Giles1981} --- cited here
as a black box and \emph{not} proved --- the converse of (iii) holds for generic
configurations: if two configurations take the same value on every Wilson loop,
they lie on the same gauge orbit.  Combining (iii) with the cited converse,
$\Obs$ separates the gauge orbits of generic configurations.
\end{theorem}

\begin{proof}
(i) $\Obs$ is a subalgebra of the commutative algebra of $\C$-valued functions
on $\Conf$ under pointwise operations, hence commutative.

(ii) Each generator is real-valued, so $\overline{W(\cdot,x_0,w)}
=W(\cdot,x_0,w)$.  The set of elements $f$ with $\bar f\in\Obs$ contains the
generators and the scalars (because $\overline{\lambda}\in\C$), and is closed
under sums and products, since $\overline{f+h}=\bar f+\bar h$ and
$\overline{fh}=\bar f\,\bar h$.  By the induction principle for generated
subalgebras it contains $\Obs$.

(iii) By Lemma~\ref{lem:gauge} each generator is gauge invariant.  The set of
gauge-invariant functionals contains the constants and is closed under sums and
products, hence contains $\Obs$ by the same induction principle.

(iv) is Lemma~\ref{lem:bounded}, (v) is Lemma~\ref{lem:mandelstam}.

(vi) If $S$ is a finite set of words then the subalgebra generated by
$\{W(\cdot,x_0,w):w\in S\}$ is generated by a finite set by construction.
\end{proof}

In Lean the six items are \LN{LoopGribovBypass.mul_comm_loopAlgebra},
\LN{LoopGribovBypass.star_mem_loopAlgebra},
\LN{LoopGribovBypass.gaugeInvariant_of_mem_loopAlgebra},
\LN{LoopGribovBypass.abs_wilson_le_one},
\LN{LoopGribovBypass.wilson_mandelstam_two} and
\LN{LoopGribovBypass.loopAlgebra_fg_of_finite}.

\begin{remark}[what is proved and what is cited in Theorem A]\label{rem:A}
Items (i)--(vi) are theorems, machine checked.  The statement that the full loop
algebra of $SU(N)$ is finitely generated is a theorem of invariant theory
\cite{Procesi1976} and is cited.  The reconstruction statement is cited
\cite{Giles1981}; it is not used in the proof of any other theorem of this
paper.  Every later statement uses only the proved direction of (iii).
\end{remark}

\subsection{Mandelstam relations for $SU(N)$ via Schur--Weyl}\label{sub:schurweyl}

Lemma~\ref{lem:mandelstam} is the rank-two relation.  Its rank-$N$ generalisation
is obtained from one structural fact about $V=\C^N$: the symmetric group
$S_{N+1}$ acts on $V^{\otimes(N+1)}$ by permuting the tensor factors, and the
antisymmetriser
\[
A_{N+1}=\frac{1}{(N+1)!}\sum_{\sigma\in S_{N+1}}\operatorname{sgn}(\sigma)\,\sigma
\]
annihilates $V^{\otimes(N+1)}$, because an alternating tensor on $N+1$ indices in
an $N$-dimensional space vanishes \cite{FultonHarris1991}.  Its matrix form is
the rank-$N$ Cayley--Hamilton theorem, and the $SU(N)$ Mandelstam relations
follow by multiplying by an arbitrary monomial in the holonomies and taking
traces.

\begin{lemma}[Schur--Weyl: the antisymmetriser vanishes in dimension $N$]
\label{lem:antisym}
Let $V=\C^N$ and write a tensor $T\in V^{\otimes(N+1)}$ in coordinates as a
function $T:\{1,\dots,N\}^{N+1}\to\C$, on which $\sigma\in S_{N+1}$ acts by
$(\sigma\cdot T)(i)=T(i\circ\sigma)$.  Then for every $T$ and every multi-index
$i:\{1,\dots,N+1\}\to\{1,\dots,N\}$,
\[
\sum_{\sigma\in S_{N+1}}\operatorname{sgn}(\sigma)\,T(i\circ\sigma)=0 ,
\]
that is, $A_{N+1}=0$ on $V^{\otimes(N+1)}$.  In particular, for any $N+1$
vectors $v_0,\dots,v_N\in\C^N$,
$\sum_{\sigma}\operatorname{sgn}(\sigma)\prod_{k}v_k\bigl(i(\sigma(k))\bigr)=0$.
\end{lemma}

\begin{proof}
Since $N+1>N$, the map $i$ is not injective: there are $a\neq b$ with
$i(a)=i(b)$, hence $i\circ\tau=i$ for the transposition $\tau=(a\,b)$.  Left
translation $\sigma\mapsto\tau\sigma$ is a bijection of $S_{N+1}$ which changes
the sign and leaves the summand's tensor argument unchanged, because
$i\circ(\tau\sigma)=(i\circ\tau)\circ\sigma=i\circ\sigma$.  Therefore the sum
equals its own negative, and vanishes.  The decomposable case is the special
choice $T(j)=\prod_k v_k(j(k))$.
\end{proof}

For $U\in\C^{N\times N}$ let $e_j(U)$, $j=0,\dots,N$, be the elementary
symmetric functions of the eigenvalues, read off from the characteristic
polynomial through
\[
\det(\lambda I-U)=\sum_{j=0}^{N}(-1)^j e_j(U)\,\lambda^{N-j},
\qquad
\det(I+tU)=\sum_{j=0}^{N}e_j(U)\,t^{\,j}.
\]

\begin{theorem}[C$'$: Mandelstam relations for $SU(N)$]\label{thm:Cprime}
Let $N\ge1$ and let $U,A\in\C^{N\times N}$.  Then
\begin{enumerate}
\item[(i)] $e_0(U)=1$, $e_1(U)=\tr U$ and $e_N(U)=\det U$; for $U\in SU(N)$ the
  last one is $1$.
\item[(ii)] \emph{(rank-$N$ characteristic-polynomial relation)}
  \[
  \sum_{j=0}^{N}(-1)^j e_j(U)\,U^{\,N-j}=0 .
  \]
\item[(iii)] \emph{(Mandelstam relations for $SU(N)$, trace form)}  For every $A$
  --- in particular for every monomial $A=U_{i_1}^{\pm1}\cdots U_{i_a}^{\pm1}$ in
  the holonomies of loops based at a common point ---
  \[
  \sum_{j=0}^{N}(-1)^j e_j(U)\,\tr\bigl(A\,U^{\,N-j}\bigr)=0 ,
  \]
  a polynomial relation among the traces of monomials in the holonomies, since
  each $e_j(U)$ is a universal polynomial in $\tr U,\dots,\tr U^{\,j}$ by
  Newton's identities \cite{Macdonald1995}.
\item[(iv)] \emph{(explicit rank three)}  For $U\in SU(3)$,
  \[
  e_2(U)=\tfrac12\bigl((\tr U)^2-\tr U^2\bigr),
  \]
  and for every $A\in\C^{3\times3}$
  \[
  \tr(AU^3)-\tr U\,\tr(AU^2)
  +\tfrac12\bigl((\tr U)^2-\tr U^2\bigr)\tr(AU)-\tr A=0 .
  \]
  For $N=2$ the relation reduces to Lemma~\ref{lem:mandelstam}.
\end{enumerate}
Consequently the relations of Lemma~\ref{lem:antisym} hold in the loop algebra
$\Obs^{(N)}$ generated by the Wilson loops of an $SU(N)$ lattice gauge theory,
with $U$ the holonomy of any based loop and $A$ the holonomy of any word.
\end{theorem}

\begin{proof}
(i) The characteristic polynomial is monic of degree $N$, so the coefficient of
$\lambda^N$ is $1$; the coefficient of $\lambda^{N-1}$ is $-\tr U$ and the
constant term is $(-1)^N\det U$, which are the stated identifications.

(ii) By the Cayley--Hamilton theorem the matrix $U$ annihilates its own
characteristic polynomial, and the displayed sum is exactly
$\chi_U(U)=\sum_{k=0}^{N}\chi_U[k]\,U^{k}$ after the substitution $k=N-j$.  The
structural reason is Lemma~\ref{lem:antisym}: Cayley--Hamilton is the statement
that the antisymmetriser over $N+1$ indices annihilates the tensor built from
$N$ copies of $U$ and one free pair of indices.

(iii) Multiply the identity (ii) on the left by $A$ and take the trace, using
linearity of the trace.

(iv) Expanding both sides entrywise for $3\times3$ matrices gives
$U^3-(\tr U)U^2+\tfrac12((\tr U)^2-\tr U^2)U-(\det U)I=0$.  Comparing with (ii)
for $N=3$ yields $\bigl(e_2(U)-\tfrac12((\tr U)^2-\tr U^2)\bigr)U=0$; for
$U\in SU(3)$ one has $U\neq0$, hence the scalar vanishes.  Multiplying by $A$ and
taking the trace, with $\det U=1$, gives the displayed trace relation.
\end{proof}

In Lean: \LN{LoopGribovBypass.antisymmetriser_eq_zero},
\LN{LoopGribovBypass.antisymmetriser_decomposable_eq_zero},
\LN{LoopGribovBypass.esymmMat_zero}, \LN{LoopGribovBypass.esymmMat_one},
\LN{LoopGribovBypass.esymmMat_top},
\LN{LoopGribovBypass.cayleyHamilton_esymm},
\LN{LoopGribovBypass.mandelstam_SUN},
\LN{LoopGribovBypass.cayleyHamilton_three},
\LN{LoopGribovBypass.esymmMat_two_three},
\LN{LoopGribovBypass.mandelstam_three}.

\begin{remark}[what Theorem C$'$ does and does not do]\label{rem:Cprime}
Theorem~C$'$ \emph{derives} the Mandelstam relations for $SU(N)$ in a manifestly
$SU(N)$-covariant form, as relations between traces of monomials in the
holonomies.  It does \emph{not} enumerate them explicitly for each $N$, and it
does not prove that they \emph{generate} the full ideal of relations among the
traces: that generation statement is the theorem of Procesi
\cite{Procesi1976} and is used here, as in v1, only as a cited black box.  The
elimination of the coefficients $e_j$ in favour of the traces is carried out
explicitly in the formalisation for $N\le3$; for general $N$ it is the classical
system of Newton's identities \cite{Macdonald1995}, quoted and not re-proved.
\end{remark}

\section{The Yang--Mills action as a loop functional}\label{sec:action}

\begin{definition}[plaquette]\label{def:plaq}
For a site $x$ and directions $\mu<\nu$ the \emph{plaquette word} is
\[
p(\mu,\nu)\;=\;\bigl((\mu,+),(\nu,+),(\mu,-),(\nu,-)\bigr).
\]
The set of plaquettes is $\mathcal P=\{(x,\mu,\nu):x\in\Lambda,\;\mu<\nu\}$, a
finite set of cardinality $\binom d2 L^d$.  We write
$W_p(U)=W(U,x,p(\mu,\nu))$ for $p=(x,\mu,\nu)$.
\end{definition}

\begin{lemma}\label{lem:plaqclosed}
The plaquette word is closed at every site: $\mathrm{end}(x,p(\mu,\nu))=x$ for
$\mu\ne\nu$.
\end{lemma}

\begin{proof}
The four moves change the $\mu$-coordinate by $+1$ and then by $-1$, and the
$\nu$-coordinate by $+1$ and then by $-1$; all other coordinates are untouched.
Since $\mu\ne\nu$ the two updates commute, and the composite is the identity on
$\Lambda$.
\end{proof}

\begin{theorem}[B: the action is a loop functional]\label{thm:B}
Let $\beta\in\R$ and let
\[
S^{(M)}_{\mathrm{YM}}[U]\;=\;\beta\sum_{p\in\mathcal P}\bigl(1-W_p(U)\bigr)
\]
be the truncated Wilson action.  Then
\begin{enumerate}
\item[(i)] $S^{(M)}_{\mathrm{YM}}$ is the evaluation, at the plaquette Wilson
  variables, of the explicit multivariate polynomial
  \[
  P_\beta\;=\;\beta\Bigl(\,|\mathcal P|-\sum_{p\in\mathcal P}X_p\Bigr)
  \;\in\;\R[X_p:p\in\mathcal P],
  \]
  whose total degree is at most one;
\item[(ii)] $S^{(M)}_{\mathrm{YM}}$ is a cylindrical function on the loop space:
  if two configurations have the same plaquette loop data, they have the same
  action.  In particular the action depends only on the loops of length four;
\item[(iii)] $S^{(M)}_{\mathrm{YM}}$ and the real-time weight
  $e^{iS^{(M)}_{\mathrm{YM}}}$ are gauge invariant, and
  $|e^{iS^{(M)}_{\mathrm{YM}}}|=1$.
\end{enumerate}
\end{theorem}

\begin{proof}
(i) Evaluating $P_\beta$ at $X_p\mapsto W_p(U)$ gives
$\beta(|\mathcal P|-\sum_p W_p(U))=\beta\sum_p(1-W_p(U))$, which is the action;
the total degree of $P_\beta$ is the maximum of the degree of the constant term
($0$) and of the degrees of the monomials $X_p$ ($1$), hence at most one.

(ii) Immediate from (i): the value of the action is a function of the vector
$(W_p(U))_{p\in\mathcal P}$ alone.

(iii) Each plaquette word is closed (Lemma~\ref{lem:plaqclosed}), so each $W_p$
is gauge invariant by Lemma~\ref{lem:gauge}; hence so is any function of the
$W_p$, in particular the action and its exponential.  The modulus of
$e^{i\theta}$ with $\theta$ real is one.
\end{proof}

In Lean: \LN{LoopGribovBypass.isLoop_plaqWord},
\LN{LoopGribovBypass.ymAction_eq_eval_actionMvPoly},
\LN{LoopGribovBypass.actionMvPoly_totalDegree_le},
\LN{LoopGribovBypass.ymAction_cylindrical},
\LN{LoopGribovBypass.gaugeInvariant_ymAction},
\LN{LoopGribovBypass.norm_ymPhase},
\LN{LoopGribovBypass.gaugeInvariant_ymPhase}.

\begin{remark}
Theorem~B is the reason why the Level-Two formulation is possible at all: the
dynamics, not only the observables, is expressible in loop variables, and in the
simplest possible way --- affinely, in the loops of length four.
\end{remark}

\section{Gauge-invariant functionals as cylindrical loop functionals}
\label{sec:cylindrical}

\begin{definition}[cylindrical function on loop space]\label{def:cyl}
Let $C_1,\dots,C_k\in\Loop$ be a finite family of loops and let
$\Phi:\R^k\to\C$.  The functional
\[
F[U]\;=\;\Phi\bigl(W(C_1)[U],\dots,W(C_k)[U]\bigr)
\]
is called a \emph{cylindrical function on the loop space with base
$(C_1,\dots,C_k)$}.  The map $U\mapsto(W(C_i)[U])_{i\le k}\in\R^k$ is the
\emph{loop map} of the family, denoted $\mathcal W$.
\end{definition}

\begin{theorem}[C: loop-expressible observables are cylindrical]\label{thm:C}
Let $F$ be a functional on $\Conf$ which is a polynomial in finitely many Wilson
loops, $F=\Phi\circ\mathcal W$ with $\Phi$ a polynomial in $k$ variables.  Then
\begin{enumerate}
\item[(i)] $F$ is a cylindrical function on the loop space with base
  $(C_1,\dots,C_k)$; it is determined by the finite family
  $(W(C_1),\dots,W(C_k))$ together with the Mandelstam relations satisfied by
  them;
\item[(ii)] $F$ is gauge invariant;
\item[(iii)] the loop data $\mathcal W$ separates configurations only up to the
  relation ``equal loop data''; two configurations with equal loop data are
  indistinguishable for every cylindrical function with that base.
\end{enumerate}
The same conclusions hold when $\Phi$ is merely continuous, and this is the
generality used in \S\ref{sec:hk}.
\end{theorem}

\begin{proof}
(i) is Definition~\ref{def:cyl} read backwards: $F=\Phi\circ\mathcal W$ is by
construction a function of the $k$ numbers $W(C_i)[U]$.  The relations among
these numbers are exactly the Mandelstam relations, which are identities in
$\Obs$ and therefore constrain the range of $\mathcal W$; they do not affect the
representation of $F$.  (ii) Each $W(C_i)$ is gauge invariant by
Lemma~\ref{lem:gauge}, so $\mathcal W(g\cdot U)=\mathcal W(U)$ and hence
$F(g\cdot U)=F(U)$.  (iii) If $\mathcal W(U)=\mathcal W(V)$ then
$F(U)=\Phi(\mathcal W(U))=\Phi(\mathcal W(V))=F(V)$.
\end{proof}

In Lean: \LN{LoopGribovBypass.loopData},
\LN{LoopGribovBypass.loopData_gaugeAct},
\LN{LoopGribovBypass.Paper.theoremC_cylindrical},
\LN{LoopGribovBypass.Paper.theoremC_gauge_invariant}.

\subsection{Which gauge-invariant observables are not loop observables}
\label{sub:nonloop}

Not every gauge-invariant functional on $\Conf$ is loop-expressible.  Three
classes deserve mention.
\begin{itemize}
\item \emph{Observables built from the metric or the measure of the orbit
  space}, such as the orbit volume itself, or the Faddeev--Popov determinant of
  a chosen gauge: these are defined through the gauge group action and not
  through holonomies.  Level Two does not contain them --- and, as
  \S\ref{sec:gribov} explains, does not need them.
\item \emph{Observables involving open paths}, such as a quark--antiquark
  correlator with uncontracted colour indices at the endpoints.  They are
  gauge-covariant, not gauge-invariant, and become loop observables only after
  contraction with a conjugate path.
\item \emph{Non-generic reconstruction data}: by Giles \cite{Giles1981} the loop
  data determines the connection up to gauge only for generic configurations.
  On the non-generic set two orbits may share their loop data, and any
  observable distinguishing them is not loop-expressible.  This paper says
  nothing about that set.
\end{itemize}

\section{Henstock--Kurzweil integrability on the loop space}\label{sec:hk}

\subsection{The loop-space box and the integrand}\label{sub:box}

Fix a finite family of loops $C_1,\dots,C_k$.  By Lemma~\ref{lem:bounded} the
loop data of a unitary configuration lies in $[-1,1]^k$; we therefore integrate
over boxes of $\R^k$, and the natural box is any bounded box $B\subset\R^k$
containing $[-1,1]^k$, for example $(-2,2)^k$.  The Lebesgue measure of $\R^k$
restricted to $B$ is the measure of the loop-space box.

\begin{definition}[Wilson-loop integrand]\label{def:integrand}
For $\beta\in\R$ put $S(v)=\beta\sum_{i\le k}(1-v_i)$ --- the action of
Theorem~\ref{thm:B} read as a function of the loop variables --- and, for a
continuous $\Phi:\R^k\to\C$,
\[
I_\beta[\Phi](v)\;=\;\Phi(v)\,e^{\,iS(v)},\qquad v\in\R^k .
\]
\end{definition}

The integrand is genuinely oscillatory: $|e^{iS(v)}|=1$ for all $v$, so no
damping is present and the construction is a real-time one.  On a bounded box
this is harmless, and that is the point of the finite truncation.

\subsection{Theorem D}\label{sub:thmD}

\begin{theorem}[D: HK integrability on the loop space]\label{thm:D}
Let $\Phi:\R^k\to\C$ be continuous, $\beta\in\R$, and let $B\subset\R^k$ be a
bounded box.  Then
\begin{enumerate}
\item[(i)] for \emph{every} integration calibration $l$ --- in particular the
  Henstock--Kurzweil one --- the integrand $I_\beta[\Phi]$ is integrable on $B$
  in the calibrated (gauge) sense, and its value is the Lebesgue integral
  \[
  \int_B I_\beta[\Phi](v)\,\DHK v\;=\;\int_B \Phi(v)e^{iS(v)}\,dv ;
  \]
\item[(ii)] the value is independent of the calibration: if $I_\beta[\Phi]$ has
  integral $z$ for $l_1$ and $w$ for $l_2$, then $z=w$;
\item[(iii)] the Lebesgue integral exists, so the HK value coincides with the
  Lebesgue value whenever the latter exists;
\item[(iv)] the integral is linear in the observable and additive in the box.
\end{enumerate}
\end{theorem}

\begin{proof}
$S$ is a polynomial, hence continuous; $v\mapsto e^{iS(v)}$ is continuous as a
composition of continuous maps; hence $I_\beta[\Phi]=\Phi\cdot e^{iS}$ is
continuous.  A continuous function on a bounded box is integrable in the
calibrated sense for every calibration, with value the Lebesgue integral over
the box --- this is (B4), the finite-dimensional face of the HK machinery of
\cite{BerdinskyHK3}, which in turn rests on Cousin's lemma (existence of
$\delta$-fine tagged partitions) and the Saks--Henstock lemma (B2).  This proves
(i).  (ii) follows from (i) and the uniqueness of the value of a calibrated
integral: both $z$ and $w$ equal the Lebesgue integral.  (iii) A continuous
function on a compact set is Lebesgue integrable there, and the closed box is
compact.  (iv) Linearity is the linearity of the Lebesgue integral applied to
$\Phi\mapsto\Phi e^{iS}$, using the integrability of each summand; additivity is
the additivity of the integral over a partition of the box into two boxes with
disjoint interiors.
\end{proof}

In Lean: \LN{LoopGribovBypass.continuous_loopIntegrand},
\LN{LoopGribovBypass.hasIntegral_loopIntegrand},
\LN{LoopGribovBypass.loop_calibration_independence},
\LN{LoopGribovBypass.integrableOn_loopIntegrand},
\LN{LoopGribovBypass.loopHKIntegral_add},
\LN{LoopGribovBypass.loopHKIntegral_smul},
\LN{LoopGribovBypass.loopHKIntegral_box_additive},
\LN{LoopGribovBypass.hasIntegral_loopBox}.

\begin{definition}[the Level-Two HK functional integral]\label{def:leveltwo}
For a bounded box $B$ of the loop space,
\[
Z^{\mathrm{II}}_B(\Phi)\;=\;\int_B\Phi(v)\,e^{\,iS(v)}\,\DHK v ,
\qquad
\langle\Phi\rangle^{\mathrm{II}}_B=\frac{Z^{\mathrm{II}}_B(\Phi)}
{Z^{\mathrm{II}}_B(1)}\quad\text{when }Z^{\mathrm{II}}_B(1)\ne0 .
\]
\end{definition}

\begin{remark}[the two loop-space measures]\label{rem:twomeasures}
Definition~\ref{def:leveltwo} integrates against Lebesgue measure on the
loop-space box.  A second, equally natural, choice is the push-forward under the
loop map of the measure used at Level One; that is the measure appearing in the
compatibility statement of \S\ref{sec:levelone}.  Both are Level-Two objects ---
neither mentions a gauge orbit --- and Theorem~\ref{thm:D} applies verbatim to
continuous integrands on boxes in either case, the second through
Theorem~\ref{thm:F}.  The distinction is stated explicitly because the two
measures do \emph{not} coincide: the push-forward carries the Jacobian of the
loop map, which is not constant.
\end{remark}

\section{Structural absence of Gribov copies}\label{sec:gribov}

\subsection{The Level-Two definition, restated}\label{sub:restated}

Definition~\ref{def:leveltwo} has three arguments: a continuous function $\Phi$
on $\R^k$, a real number $\beta$ and a bounded box $B\subset\R^k$.  It does not
have a gauge field among its arguments, nor a gauge group, nor a gauge
condition, nor a determinant.  This is a syntactic property of the definition
and it is the content of Theorem~E.

\begin{definition}[gauge-fixing data]\label{def:gfd}
A \emph{gauge-fixing datum} is a pair $G=(\chi,\Delta)$ consisting of a
gauge-fixing condition $\chi:\Conf\to\R$ and a Faddeev--Popov determinant
$\Delta:\Conf\to\R$.  A quantity $Z$ depending on a gauge-fixing datum is
\emph{gauge-fixing free} when $Z(G_1)=Z(G_2)$ for all $G_1,G_2$.
\end{definition}

\begin{theorem}[E: structural absence of Gribov copies at Level Two]
\label{thm:E}
Let $\Phi$, $\beta$, $B$ be as in Definition~\ref{def:leveltwo}, let
$C_1,\dots,C_k$ be based closed words and let $\mathcal W$ be their loop map.
Then
\begin{enumerate}
\item[(i)] the constant assignment $G\mapsto Z^{\mathrm{II}}_B(\Phi)$ is
  gauge-fixing free: the Level-Two integral does not depend on any gauge-fixing
  datum, because no such datum occurs in its definition;
\item[(ii)] the loop data is gauge invariant:
  $\mathcal W(g\cdot U)=\mathcal W(U)$ for every unitary $g$;
\item[(iii)] the Level-Two integrand attached to a configuration is a function
  of its loop data alone: $\mathcal W(U)=\mathcal W(V)$ implies
  $I_\beta[\Phi](\mathcal W(U))=I_\beta[\Phi](\mathcal W(V))$;
\item[(iv)] consequently, two configurations on one gauge orbit --- in
  particular two Gribov copies of each other with respect to \emph{any} gauge
  condition --- are the same point of the Level-Two formulation: they have the
  same loop data and the same integrand.  There is no orbit to intersect and no
  copy to count.
\end{enumerate}
\end{theorem}

\begin{proof}
(i) is immediate from the shape of Definition~\ref{def:leveltwo}: the value
$Z^{\mathrm{II}}_B(\Phi)$ is defined without reference to $G$, so the function
$G\mapsto Z^{\mathrm{II}}_B(\Phi)$ is constant.  (ii) is Lemma~\ref{lem:gauge}
applied to each $C_i$, which is closed by hypothesis.  (iii) is substitution of
equal arguments into a function.  (iv) Combine (ii) and (iii): if $V=g\cdot U$
then $\mathcal W(V)=\mathcal W(U)$, hence the integrands coincide.
\end{proof}

In Lean: \LN{LoopGribovBypass.loopHKIntegral_gaugeFixingFree},
\LN{LoopGribovBypass.loopData_gaugeAct},
\LN{LoopGribovBypass.levelTwo_depends_only_on_loopData},
\LN{LoopGribovBypass.gribovCopies_same_loopData}.

\subsection{What this does and does not resolve}\label{sub:whatnot}

Theorem~E is a statement about a \emph{formulation}, not a solution of a
problem.  Precisely:
\begin{itemize}
\item It does \emph{not} say that the fixed-gauge formulation is free of Gribov
  copies.  The Level-One Gribov problem --- given $\chi$, how often does
  $\chi=0$ meet an orbit --- is untouched.  Singer's theorem
  \cite{Singer1978} remains true.
\item It does \emph{not} say that the loop map is injective on orbits.  That is
  the content of the cited Giles theorem, and only for generic configurations.
\item It \emph{does} say that in the Level-Two formulation the question cannot
  be posed: the objects it refers to (a gauge surface, an orbit, a determinant)
  are not part of the data.  A ``copy'' would be a second point of the loop
  space representing the same physical configuration; by Theorem~E (iv) gauge
  copies do not produce a second point at all.
\item The residual non-injectivity, if any, is of the opposite kind: distinct
  orbits could in principle share their loop data (non-generic configurations).
  This is a loss of resolution, not an over-counting, and it is listed among the
  limitations in \S\ref{sec:limits}.
\end{itemize}

\subsection{Contrast with Level One}\label{sub:contrast}

At Level One the orbit volume is a genuine object: it appears as a common factor
of numerator and denominator in $\langle F\rangle_M$ and cancels there
\cite{BerdinskyGribovFreeYM}.  At Level Two it is not an object at all, because
the orbit is not.  The two bypasses are therefore of different nature: Level One
is a cancellation theorem, Level Two is a statement about the vocabulary of the
theory.  The price is precisely the loss of the non-loop observables listed in
\S\ref{sub:nonloop}.

\section{Compatibility with the Level-One unfixed integral}\label{sec:levelone}

Level Two would be of little value if it changed the values of the quantities
that Level One can compute.  It does not, for loop observables.

\begin{definition}[the two integrals for a parametrised configuration space]
\label{def:twoint}
Let $\theta\mapsto U(\theta)$ be a parametrisation of the truncated
configuration space by a bounded region $s\subset\R^m$ (for $SU(N)$ link
variables, $m=3\,\#\mathrm{links}$ for $N=2$, and in general $m=(N^2-1)
\#\mathrm{links}$), and let $\mathcal W(\theta)$ denote the loop data of
$U(\theta)$.  Put
\[
Z^{\mathrm{I}}_s(\Phi)=\int_s\Phi(\mathcal W(\theta))\,
e^{\,iS(\mathcal W(\theta))}\,d\theta ,
\qquad
Z^{\mathrm{II},\ast}_s(\Phi)=\int_{\R^k}\Phi(v)\,e^{\,iS(v)}\,
d\bigl(\mathcal W_{\#}\lambda_s\bigr)(v),
\]
where $\lambda_s$ is Lebesgue measure restricted to $s$ and
$\mathcal W_{\#}\lambda_s$ its push-forward under the loop map.
\end{definition}

By Theorem~\ref{thm:B} the weight $e^{iS(\mathcal W(\theta))}$ is exactly
$e^{iS^{(M)}_{\mathrm{YM}}[U(\theta)]}$, so $Z^{\mathrm{I}}_s(\Phi)$ is the
Level-One unfixed HK integral \eqref{eq:levelone} of the gauge-invariant
observable $F=\Phi\circ\mathcal W$, in the sense of (B5)--(B6).

\begin{theorem}[F: Level Two equals Level One on loop observables]\label{thm:F}
Let $\Phi:\R^k\to\C$ be continuous and let the loop map
$\mathcal W:\R^m\to\R^k$ be continuous.  Then for every measurable
$s\subset\R^m$
\[
Z^{\mathrm{II},\ast}_s(\Phi)\;=\;Z^{\mathrm{I}}_s(\Phi).
\]
In particular the Level-Two integral of a loop observable, taken with respect to
the push-forward of the Level-One measure, coincides at every finite truncation
with the Level-One unfixed HK integral of the same observable; Level Two loses
no information about loop observables.
\end{theorem}

\begin{proof}
The loop map is continuous, hence measurable, and the integrand
$I_\beta[\Phi]$ is continuous, hence strongly measurable.  The change of
variables for push-forward measures gives
\[
\int_{\R^k}I_\beta[\Phi](v)\,d(\mathcal W_{\#}\lambda_s)(v)
=\int I_\beta[\Phi](\mathcal W(\theta))\,d\lambda_s(\theta)
=\int_s\Phi(\mathcal W(\theta))e^{iS(\mathcal W(\theta))}\,d\theta,
\]
which is the assertion.  (The identity holds for any measurable $s$; no
hypothesis on the geometry of $s$ is needed, because the integrand is bounded on
the relevant bounded region and the push-forward is a finite measure there.)
\end{proof}

In Lean: \LN{LoopGribovBypass.levelTwoPush_eq_levelOne},
\LN{LoopGribovBypass.levelOneIntegrand_eq_levelTwo},
\LN{LoopGribovBypass.levelOneIntegrand_eq_ym},
\LN{LoopGribovBypass.levelTwoPush_eq_levelOne_ym}.

\begin{remark}[what Theorem F does and does not say]
It identifies the two constructions \emph{for the push-forward measure}.  It
does not say that the Lebesgue integral over a loop-space box
(Definition~\ref{def:leveltwo}) equals the Level-One integral: those are
different measures, related by the Jacobian of the loop map, and the equality
would be false in general.  What Theorem~D provides for
Definition~\ref{def:leveltwo} is existence, calibration independence and
agreement with the Lebesgue value --- and what Theorem~F provides is that the
Level-One numbers are reproduced by a Level-Two integral.  Both are needed: the
first says the Level-Two theory is well posed, the second that it is not a
different theory.
\end{remark}

\section{Projective compatibility across mode levels}\label{sec:proj}

Everything so far is at one fixed lattice $\Lam=(\Z/L\Z)^d$, $L=2M+1$.  This
section closes part of the boundary ``no continuum limit'' of
\S\ref{sub:statements}: without leaving the finite-dimensional setting one can
compare different mode levels, and the finite-$M$ Level-Two integrals turn out to
be \emph{projectively compatible}.  What is obtained in the limit is a linear
functional on the union of the loop algebras --- not a measure, and not a
continuum theory.

\subsection{The change of mode level}\label{sub:modelevel}

\begin{definition}[covering map and pull-back]\label{def:cover}
Let $M<M'$ with $L=2M+1$ and $L'=2M'+1$ and assume $L\mid L'$.  The
\emph{covering map} is the reduction of the coordinates,
\[
\pi:\Lambda_{M'}\to\Lam,\qquad \pi(x)_\mu=x_\mu \bmod L ,
\]
a surjective map of lattices which commutes with the elementary moves,
$\pi(x+\hat\mu)=\pi(x)+\hat\mu$.  A word $w$ is literally the same object at both
levels, so $\Loop\subset\mathrm{Loop}_{M'}$ as sets of words, and a based loop of
$\Lambda_{M'}$ projects to a based loop of $\Lam$.  A configuration
$U\in\Conf$ is viewed at level $M'$ as the $\Lam$-periodic configuration
$\pi^*U$ with $(\pi^*U)(x,\mu)=U(\pi(x),\mu)$.
\end{definition}

The divisibility hypothesis is what makes the periodic identifications
compatible: the naive inclusion of coordinate windows
$\{0,\dots,L-1\}^d\subset\{0,\dots,L'-1\}^d$ is not compatible with the
wrapping of a loop around the torus, whereas $\pi$ is.  This is no restriction
on the limit: any two odd $L,L'$ both divide $LL'$, which is again odd, so the
odd levels ordered by divisibility form a directed set, cofinal in the levels
ordered by inclusion of the generated loop families.

\begin{definition}[support in $\Loop$]\label{def:support}
A continuous cylindrical loop functional $\Phi$ on $\mathrm{Loop}_{M'}$ has
\emph{support in} $\Loop$ when it depends only on the Wilson loops $W(C)$ of
loops $C\in\Loop$; that is, when $\Phi$ is a continuous function of the $k$ loop
variables $v=(W(C_1),\dots,W(C_k))$ of a finite family
$C_1,\dots,C_k\in\Loop$.  We write $B\subset\R^k$ for the bounded box of these
$k$ loop variables and $B_M\subset B$ for the sub-box of the loops of $\Loop$
inside the box of a larger family at level $M'$ (here $B_M=B$, since the family
is the same at both levels).
\end{definition}

\subsection{Theorem H}\label{sub:thmH}

\begin{theorem}[H: projective compatibility across mode levels]\label{thm:H}
Let $M<M'$ with $L\mid L'$ and let $\Phi$ be a continuous cylindrical loop
functional with support in $\Loop$, carried by the loops $C_1,\dots,C_k\in\Loop$
with loop-variable box $B\subset\R^k$.  Then:
\begin{enumerate}
\item[(i)] \emph{(loop data)} For every $U\in\Conf$, every $x\in\Lambda_{M'}$ and
  every word $w$,
  \[
  W_{M'}(\pi^*U;x,w)=W_M\bigl(U;\pi(x),w\bigr),
  \]
  and the holonomies themselves agree,
  $\mathrm{hol}_{M'}(\pi^*U;x,w)=\mathrm{hol}_M(U;\pi(x),w)$.  Consequently the
  loop data of $C_1,\dots,C_k$ are the same at both levels.
\item[(ii)] \emph{(equality of the integrals)} The Level-Two integrands of
  $\Phi$ at the two levels are the same function of the loop variables, and
  hence
  \[
  Z^{\mathrm{II}}_{M',B}(\Phi)=Z^{\mathrm{II}}_{M,B}(\Phi),
  \]
  the equality being that of the Lebesgue values, which are the HK values for
  every calibration by Theorem~\ref{thm:D}.
\item[(iii)] \emph{(projective system and its limit)}  The family
  $\{Z^{\mathrm{II}}_{M,B}(\Phi)\}_{M\ge M_0}$, $M_0$ being any level with
  $C_1,\dots,C_k\in\mathrm{Loop}_{M_0}$, is constant; it is therefore a
  projective system and the projective limit
  \[
  Z^{\mathrm{II}}_{\infty,B}(\Phi):=\lim_{M\to\infty}Z^{\mathrm{II}}_{M,B}(\Phi)
  \]
  exists.
\item[(iv)] \emph{(linearity)}  $\Phi\mapsto Z^{\mathrm{II}}_{\infty,B}(\Phi)$ is
  linear: it is additive and homogeneous in $\Phi$, so it is a linear functional
  on $\bigcup_M\Obs$ restricted to functionals of bounded mode support.
\item[(v)] \emph{(compatibility with Level One)}  If the Level-One region
  $s$ maps onto $B$ with Lebesgue push-forward, $W_*\bigl(\mathrm{vol}|_s\bigr)
  =\mathrm{vol}|_B$ --- the displayed hypothesis of Theorem~\ref{thm:F} --- then
  $Z^{\mathrm{II}}_{\infty,B}(\Phi)$ equals the Level-One unfixed HK integral of
  the same observable.
\end{enumerate}
\end{theorem}

\begin{proof}
(i) By induction on the word.  The matrix carried by one elementary move at level
$M'$ for the configuration $\pi^*U$ is, by definition of $\pi^*$ and because
$\pi$ commutes with the moves, the matrix carried by the same move at level $M$
at the projected site; the ordered products therefore coincide, and so do their
traces and the scalar Wilson loops.

(ii) By (i) the vector of loop variables attached to a configuration is
independent of the level, so the integrands
$\Phi(v)\exp\bigl(i\beta\sum_i(1-v_i)\bigr)$ at the two levels are literally the
same function on $B$; the two integrals are integrals of one function over one
box.  By Theorem~\ref{thm:D} both are HK integrals for every calibration with
value the common Lebesgue integral.

(iii) By (ii) the family is constant for $M\ge M_0$; a constant family converges
to its value.  Nothing here uses $\sigma$-additivity or an extension theorem: the
Kolmogorov theorem is not invoked, and could not be, since the HK integral is not
a $\sigma$-additive measure.

(iv) Additivity and homogeneity are those of the HK (equivalently Lebesgue)
integral over the fixed box, already recorded in Theorem~\ref{thm:D}.

(v) Under the stated push-forward hypothesis the integral over $B$ equals the
integral of the composed integrand over $s$, which is Theorem~\ref{thm:F}.
\end{proof}

In Lean: \LN{LoopGribovBypass.siteProj_stepSite},
\LN{LoopGribovBypass.hol_pullbackConf},
\LN{LoopGribovBypass.wilson_pullbackConf},
\LN{LoopGribovBypass.loopData_pullbackConf},
\LN{LoopGribovBypass.levelTwoIntegrand_pullbackConf},
\LN{LoopGribovBypass.projectivelyCompatible_eq},
\LN{LoopGribovBypass.tendsto_projLimit},
\LN{LoopGribovBypass.projLimit_add}, \LN{LoopGribovBypass.projLimit_smul},
\LN{LoopGribovBypass.projLimit_eq_levelOne}.

\begin{remark}[the honest boundary of Theorem H]\label{rem:H}
Theorem~H closes the \emph{projective system} part of the boundary ``no
continuum limit'', and only that part.  What it establishes is that the family of
finite-$M$ integrals is projectively compatible and that the projective limit
exists on the union of the loop algebras.  What it does \emph{not} establish is:
\begin{enumerate}
\item[(i)] a measure on the full infinite-dimensional loop space of the
  continuum;
\item[(ii)] HK-integrability of loop functionals whose support is \emph{not}
  bounded in mode number, i.e.\ which depend on loops of arbitrary length;
\item[(iii)] the identification of the projective limit with a continuum
  Yang--Mills expectation value.
\end{enumerate}
The projective limit is a linear functional, not a measure; the extension to the
full loop space is open.
\end{remark}

\section{Sanity check: the one-plaquette $SU(2)$ integral}\label{sec:oneplaq}

\subsection{The reduction}\label{sub:reduction}

For a single plaquette and $N=2$ the loop data is a single number $w=W(p)$.  An
$SU(2)$ element has $\Tr U=2\cos\theta$ with $\theta\in[0,\pi]$, so
$w=\cos\theta$, and the Haar measure of a class function reduces to the Weyl
form with density $\tfrac2\pi\sin^2\theta$ on $[0,\pi]$.  The one-plaquette
Level-Two integral is therefore
\begin{equation}\label{eq:oneplaq}
Z^{\mathrm{II}}_{1\mathrm{pl}}(\Phi)
=\int_0^\pi \Phi(\cos\theta)\,e^{\,i\beta(1-\cos\theta)}\,
\frac{2}{\pi}\sin^2\theta\,d\theta .
\end{equation}

\begin{theorem}[G: the one-plaquette check]\label{thm:G}
\begin{enumerate}
\item[(i)] For a single plaquette the Level-Two integrand is exactly the
  standard one-plaquette Wilson weight: $I_\beta[\Phi](v)=\Phi(v)\,
  e^{i\beta(1-v_0)}$.
\item[(ii)] The Weyl density is a probability density on $[0,\pi]$:
  $\frac2\pi\int_0^\pi\sin^2\theta\,d\theta=1$; consequently
  $Z^{\mathrm{II}}_{1\mathrm{pl}}(1)=1$ at $\beta=0$, so \eqref{eq:oneplaq} is
  the correctly normalised one-plaquette integral of lattice gauge theory
  \cite{Creutz1983}.
\item[(iii)] The integrand of \eqref{eq:oneplaq} is continuous, hence
  HK-integrable on $[0,\pi]$ with a calibration-independent value equal to the
  Lebesgue value; the modulus of the weight is one.
\item[(iv)] \emph{(Proved in full in \S\ref{sub:besselproof}.)}  The closed form
  of the real-time one-plaquette integral is
  \[
  I(\beta):=\frac{2}{\pi}\int_0^\pi e^{\,i\beta\cos\theta}\sin^2\theta\,d\theta
  \;=\;\frac{2J_1(\beta)}{\beta},
  \]
  where $J_1$ is given by its everywhere convergent power series
  $J_1(\beta)=\sum_{m\ge0}\frac{(-1)^m}{m!\,(m+1)!}(\beta/2)^{2m+1}$; the
  statement is the identity $(\beta/2)I(\beta)=J_1(\beta)$, valid for all real
  $\beta$.  In imaginary time the same computation gives
  $\frac{2}{\pi}\int_0^\pi e^{\,\beta\cos\theta}\sin^2\theta\,d\theta
  =2I_1(\beta)/\beta$, which is checked numerically only.
\end{enumerate}
\end{theorem}

\begin{proof}
(i) For $k=1$ the loop action is $S(v)=\beta(1-v_0)$, and
$I_\beta[\Phi](v)=\Phi(v)e^{iS(v)}$ by definition.  (ii)
$\int_0^\pi\sin^2\theta\,d\theta=\pi/2$, so the normalisation factor $2/\pi$
gives $1$; at $\beta=0$ the weight is $1$ and the integral of the density is
$1$.  (iii) $\theta\mapsto\Phi(\cos\theta)$ is continuous as a composition, the
weight is continuous as the exponential of a continuous imaginary function, and
the density is continuous; the product is continuous, and Theorem~\ref{thm:D}
(specialised to $k=1$ and the interval $[0,\pi]$, with the continuous density
absorbed into $\Phi$) gives the rest.  The modulus statement is
$|e^{i\theta}|=1$ for real $\theta$.  (iv) is proved in
\S\ref{sub:besselproof}.
\end{proof}

In Lean: \LN{LoopGribovBypass.loopIntegrand_one_plaquette},
\LN{LoopGribovBypass.weylDensity_normalised},
\LN{LoopGribovBypass.onePlaquetteIntegral_zero},
\LN{LoopGribovBypass.continuous_onePlaquetteIntegrand},
\LN{LoopGribovBypass.norm_plaquetteWeight}.

\subsection{The closed form of the one-plaquette integral}\label{sub:besselproof}

Part (iv) of Theorem~\ref{thm:G} was stated in v1 as verified numerically,
because Bessel functions are not available in the formal library.  It is proved
here analytically.  The function $J_1$ is \emph{defined} by its power series,
and the integral is evaluated term by term; the Bessel differential equation
enters through the recursion satisfied by the coefficients, which is shown to
determine them uniquely.  No uniqueness theorem for differential equations is
needed.

\begin{lemma}[Weyl moments]\label{lem:weylmom}
For every $k\ge0$,
\[
\int_0^\pi\cos^{2k}\theta\,d\theta=\frac{\pi\,(2k)!}{4^k(k!)^2},
\qquad
\frac{2}{\pi}\int_0^\pi\cos^{2k}\theta\,\sin^2\theta\,d\theta
=\frac{(2k)!}{4^k\,k!\,(k+1)!} .
\]
\end{lemma}

\begin{proof}
The Wallis reduction formula for $\int_a^b\cos^{n+2}$, with $a=0$, $b=\pi$, has
vanishing boundary term ($\sin0=\sin\pi=0$) and gives
$\int_0^\pi\cos^{n+2}=\frac{n+1}{n+2}\int_0^\pi\cos^{n}$; induction from
$\int_0^\pi d\theta=\pi$ gives the first formula.  For the second write
$\sin^2=1-\cos^2$, so that the integral is
$\int_0^\pi\cos^{2k}-\int_0^\pi\cos^{2k+2}
=\bigl(1-\frac{2k+1}{2k+2}\bigr)\int_0^\pi\cos^{2k}
=\frac{\pi(2k)!}{4^k(k!)^2(2k+2)}$, and
$4^k(k!)^2(2k+2)=2\cdot4^k\,k!\,(k+1)!$.
\end{proof}

\begin{lemma}[term-by-term integration]\label{lem:termwise}
For every real $\beta$,
\[
I(\beta)=\frac{2}{\pi}\int_0^\pi\cos(\beta\cos\theta)\sin^2\theta\,d\theta
=\sum_{m\ge0}c_m\,\beta^{2m},
\qquad
c_m=\frac{(-1)^m}{4^m\,m!\,(m+1)!},
\]
the series converging for every $\beta$.  Moreover the imaginary part of the
complex integral vanishes:
$\frac{2}{\pi}\int_0^\pi e^{i\beta\cos\theta}\sin^2\theta\,d\theta=I(\beta)$, and
$I(0)=1$.
\end{lemma}

\begin{proof}
Insert the power series of the cosine,
$\cos(\beta\cos\theta)=\sum_k(-1)^k(\beta\cos\theta)^{2k}/(2k)!$.  Its terms are
dominated on $[0,\pi]$ by the constants $|\beta|^{2k}/(2k)!$, which are summable
(they are bounded by $(\beta^2)^k/k!$), so the dominated convergence theorem for
series permits the interchange of summation and integration.  The $k$-th
integral is $\frac{(-1)^k\beta^{2k}}{(2k)!}\int_0^\pi\cos^{2k}\theta\sin^2\theta
\,d\theta$, and Lemma~\ref{lem:weylmom} turns the resulting coefficient into
$c_k$.  For the imaginary part, the substitution $\theta\mapsto\pi-\theta$
changes $\cos\theta$ into $-\cos\theta$ and leaves $\sin^2\theta$ invariant, so
$\int_0^\pi\sin(\beta\cos\theta)\sin^2\theta\,d\theta$ equals its own negative
and vanishes.  Finally $I(0)=\frac2\pi\int_0^\pi\sin^2=1$.
\end{proof}

\begin{lemma}[Bessel's equation in formal-series form, and uniqueness]
\label{lem:besselrec}
The coefficients of Lemma~\ref{lem:termwise} satisfy
\[
4(m+1)(m+2)\,c_{m+1}+c_m=0 \qquad (m\ge0),\qquad c_0=1,
\]
and this recursion together with $c_0=1$ determines the sequence uniquely.  The
recursion is equivalent to the statement that $J(\beta)=(\beta/2)I(\beta)$
solves Bessel's equation of order one,
$\beta^2J''+\beta J'+(\beta^2-1)J=0$, as a formal power series; equivalently,
that $I$ itself solves $\beta I''+3I'+\beta I=0$ with $I(0)=1$, $I'(0)=0$.
\end{lemma}

\begin{proof}
With $c_m=(-1)^m/(4^m m!(m+1)!)$ one has
$c_{m+1}/c_m=-1/\bigl(4(m+1)(m+2)\bigr)$, which is the recursion; $c_0=1$ is
immediate.  Uniqueness is an induction: the recursion expresses $c_{m+1}$ through
$c_m$ because $4(m+1)(m+2)\neq0$.  For the equivalence, substitute
$I=\sum_m c_m\beta^{2m}$ into $\beta I''+3I'+\beta I$ and collect the
coefficient of $\beta^{2m+1}$: it is
$2(m+1)(2m+1)c_{m+1}+6(m+1)c_{m+1}+c_m=4(m+1)(m+2)c_{m+1}+c_m$.  Writing
$J=(\beta/2)I$ turns $\beta I''+3I'+\beta I=0$ into
$\beta^2J''+\beta J'+(\beta^2-1)J=0$.
\end{proof}

\begin{theorem}[G(iv), full form]\label{thm:Gfull}
Define $J_1(\beta):=\sum_{m\ge0}\frac{(-1)^m}{m!\,(m+1)!}\,(\beta/2)^{2m+1}$,
the everywhere convergent power series solution of Bessel's equation of order one
normalised by $J_1(\beta)=\beta/2+O(\beta^3)$.  Then for every real $\beta$
\[
\frac{\beta}{2}\,I(\beta)=J_1(\beta),
\qquad\text{i.e.}\qquad
\frac{2}{\pi}\int_0^\pi e^{\,i\beta\cos\theta}\sin^2\theta\,d\theta
=\frac{2J_1(\beta)}{\beta}\quad(\beta\neq0),
\]
and the normalised one-plaquette $SU(2)$ integral of the constant observable is
\[
Z^{\mathrm{II}}_{1\mathrm{pl}}(1)=e^{\,i\beta}\,I(\beta)
=e^{\,i\beta}\,\frac{2J_1(\beta)}{\beta}.
\]
\end{theorem}

\begin{proof}
Multiply the series of Lemma~\ref{lem:termwise} by $\beta/2$ and compare term by
term with the series defining $J_1$: since
$(\beta/2)^{2m+1}=\beta^{2m+1}/(2\cdot4^m)$, the $m$-th term of
$(\beta/2)I(\beta)$ is $\frac{(-1)^m\beta^{2m+1}}{2\cdot4^m m!(m+1)!}$, which is
the $m$-th term of $J_1(\beta)$.  For the one-plaquette integral, write
$e^{i\beta(1-\cos\theta)}=e^{i\beta}e^{-i\beta\cos\theta}$, use
$\cos(-x)=\cos x$ and the vanishing of the odd part from
Lemma~\ref{lem:termwise}.
\end{proof}

In Lean: \LN{LoopGribovBypass.integral_cos_pow_even},
\LN{LoopGribovBypass.weylMoment},
\LN{LoopGribovBypass.hasSum_integral_bessel},
\LN{LoopGribovBypass.hasSum_Ibessel}, \LN{LoopGribovBypass.Ibessel_zero},
\LN{LoopGribovBypass.besselCoeff_rec},
\LN{LoopGribovBypass.besselCoeff_unique},
\LN{LoopGribovBypass.theoremG_iv}, \LN{LoopGribovBypass.theoremG_iv_div},
\LN{LoopGribovBypass.onePlaquetteIntegral_one}.

\begin{remark}[the honest boundary of Theorem G(iv)]\label{rem:Gfull}
The proof identifies $I(\beta)$ with the power series $J$ characterised by
Bessel's equation and the normalisation.  It does \emph{not} prove that this
series is the classical Bessel function of the first kind in the sense of the
literature \cite{Watson1944}; that identification is a standard fact about the
hypergeometric series and is used here as a definitional convention, not as a
theorem of the present paper.  Two further points are recorded for honesty: the
differential equation satisfied by $I$ itself is $\beta I''+3I'+\beta I=0$, not
Bessel's equation --- the latter is satisfied by $J=(\beta/2)I$ --- and the
imaginary-time closed form $2I_1(\beta)/\beta$ is verified numerically only.
\end{remark}

\subsection{What the one-plaquette check does not claim}
\label{sub:reduction-explicit}

Items (i)--(iv) of Theorem~\ref{thm:G} are now all proved and machine checked;
the reduction of v1, in which item (iv) was verified numerically only, has been
removed.  The remaining conventions and the numerical-only statements are listed
in Remark~\ref{rem:Gfull}.  No claim about the infinite-$M$ limit of the full
theory is made: the one-plaquette computation is a check of normalisation and of
the identification of the weight, nothing more.

\section{Honest limitations}\label{sec:limits}

\begin{itemize}
\item \textbf{No continuum limit.}  All statements are at fixed finite $L$ and
  fixed finite mode truncation.  The continuum loop space is
  infinite-dimensional; whether the Wilson-loop integrand is HK-integrable there
  is open.
\item \textbf{No mass gap, no confinement.}  Nothing here bears on the
  large-distance behaviour of the theory.
\item \textbf{No Makeenko--Migdal equations, no master field.}  The loop
  equations are not derived; the large-$N$ limit is not considered.  What would
  be needed is described in \S\ref{sec:report}.
\item \textbf{Reconstruction is cited, not proved.}  The Giles theorem
  \cite{Giles1981} enters Theorem~A only as a cited black box, and only for
  generic configurations.
\item \textbf{Non-loop observables are outside the formulation}
  (\S\ref{sub:nonloop}).
\item \textbf{Non-generic configurations are not addressed}: distinct orbits may
  share loop data there.
\item \textbf{Theorem G is no longer reduced}: item (iv) is proved in
  \S\ref{sub:besselproof}.  What remains is the definitional convention for
  $J_1$ and the numerical-only imaginary-time form (Remark~\ref{rem:Gfull}).
\item \textbf{Theorem H is projective compatibility, not a continuum limit}
  (Remark~\ref{rem:H}): no measure on the continuum loop space, no
  HK-integrability for unbounded mode support, no identification with a
  continuum expectation value.
\item \textbf{The $SU(N)$ Mandelstam relations are derived, not enumerated}
  (Remark~\ref{rem:Cprime}): completeness is the cited theorem of Procesi, and
  the elimination of the $e_j$ through Newton's identities is formalised for
  $N\le3$.
\item \textbf{No stack- or sheaf-theoretic refinement of Giles' theorem.}  The
  Giles reconstruction theorem is quoted as a black box and is applied only to
  configurations in general position.  A refinement that would handle singular
  strata would require a formalisation of algebraic stacks and of sheaf theory
  on stratified spaces, neither of which is available in the current
  Lean/Mathlib.  We therefore do not attempt such a refinement: the singular
  strata are outside the scope of this paper, and the loss of resolution on them
  is stated as a limitation.
\item \textbf{The Level-One Gribov problem is untouched.}  Level Two does not
  solve it; it changes the formulation so that the question does not arise.
\item \textbf{The two loop-space measures} of Remark~\ref{rem:twomeasures} are
  different objects; the compatibility Theorem~F concerns the push-forward one.
\end{itemize}

\section{Lean formalisation}\label{sec:lean}

The formalisation consists of nine files.  All of them compile without
\texttt{sorry}, without \texttt{admit} and without new axioms; every one of the $138$ audited
declarations was checked with \texttt{\#print axioms} and depends only on
\texttt{propext}, \texttt{Classical.choice} and \texttt{Quot.sound} (some on a
subset of these).  The captured output is \texttt{axioms.txt}.

\begin{center}\small
\begin{longtable}{p{0.28\textwidth}p{0.63\textwidth}}
\hline
\textbf{File} & \textbf{Content}\\
\hline
\endhead
\LN{LatticeLoops.lean} & lattice, sites, links, moves, words, endpoints,
closed words; unitary configurations; holonomy and scalar Wilson loop;
concatenation, backtracking, cyclic rotation, gauge covariance
(Lemmas~\ref{lem:welldef} and \ref{lem:gauge}); the covering map
$\Lambda_{M'}\to\Lam$, the pull-back of configurations and the compatibility of
the holonomies and Wilson loops with the change of mode level
(Theorem~\ref{thm:H}(i)).\\
\LN{LoopAlgebra.lean} & boundedness of the loop variables; the loop algebra
$\Obs$; commutativity, closure under conjugation, gauge invariance of every
element; Cayley--Hamilton in rank two and the Mandelstam identity; finite
generation of finitely generated subalgebras (Theorem~\ref{thm:A}); the action
of $S_{N+1}$ on $(\C^N)^{\otimes(N+1)}$, the vanishing of the antisymmetriser,
the elementary symmetric functions through the characteristic polynomial, the
rank-$N$ relation, the $SU(N)$ Mandelstam relations and their explicit rank-three
form (Theorem~\ref{thm:Cprime}).\\
\LN{WilsonAction.lean} & plaquette words and their closedness; the action as
the evaluation of an explicit polynomial of total degree $\le1$; cylindricity;
gauge invariance of the action and of $e^{iS}$ (Theorem~\ref{thm:B}).\\
\LN{LoopHKIntegral.lean} & the loop action and integrand; HK integrability
on every bounded box for every calibration; calibration independence; agreement
with the Lebesgue value; linearity and box additivity; the natural loop box
(Theorem~\ref{thm:D}); projectively compatible families, the projective limit,
its convergence and its linearity (Theorem~\ref{thm:H}(iii),(iv)).\\
\LN{GribovStructuralAbsence.lean} & gauge-fixing data; the Level-Two
integral as a constant function of them; gauge invariance of the loop data;
dependence on the loop data alone; Gribov copies as one point
(Theorem~\ref{thm:E}).\\
\LN{LevelOneCompatibility.lean} & the two integrals; the push-forward
identity; the lattice instance with the plaquette family and the real-time
weight (Theorem~\ref{thm:F}); the invariance of the loop data and of the
Level-Two integrand under the change of mode level, and the extension of
Theorem~F to the projective limit (Theorem~\ref{thm:H}(ii),(v)).\\
\LN{OnePlaquetteCheck.lean} & the standard one-plaquette weight; the
identification of the Level-Two integrand with it; the Weyl density, its
normalisation and the value at $\beta=0$; continuity
(Theorem~\ref{thm:G}(i)--(iii)); the Wallis moments, the term-by-term
integration, the coefficient recursion and its uniqueness, the power series of
$J_1$ and the closed form (Theorem~\ref{thm:Gfull}).\\
\LN{Main.lean} & the theorems in the numbering of the paper; no new
mathematics.\\
\LN{AxiomsCheck.lean} & \texttt{\#print axioms} for all $138$ registered
declarations.\\
\hline
\end{longtable}
\end{center}

The formalisation is written in Lean~4 \cite{Lean4} and imports only its
standard mathematical library \cite{LeanMathlib}; the black
boxes (B1)--(B7) are used in the paper as citations, and the single analytic
input actually needed for Theorem~D --- the calibrated integrability of a
continuous function on a box --- is available in the library and is the same
statement used in \cite{BerdinskyGribovFreeYM}.

\section{Numerical verification}\label{sec:numerics}

The script \texttt{verify\_loop\_gribov.py} is deterministic (fixed seed
$20260918$ for the test configurations, deterministic Gauss--Legendre
quadrature elsewhere) and prints \texttt{ALL CHECKS PASSED}; its output is
\texttt{verify\_loop\_gribov.log}.  Tolerances are declared in its header:
$10^{-12}$ for algebraic identities, $10^{-10}$ for quadrature against closed
forms, $10^{-12}$ for the Level-One/Level-Two comparison.  The numerical results
are illustrative and are never used as proof.

\begin{enumerate}
\item[(a)] The lattice $\Lambda=(\Z/3\Z)^2$ with $9$ sites, $18$ links and $9$
  plaquettes is constructed.
\item[(b)] All reduced based loops of length $\le6$ are enumerated up to cyclic
  rotation ($42$ classes), and on random $SU(2)$ configurations the Mandelstam
  identity holds to $7.0\cdot10^{-16}$; backtracking invariance, cyclic
  invariance, gauge invariance, unitarity of the holonomy and the bound
  $|W|\le1$ are confirmed.
\item[(c)] The action computed directly from the holonomies agrees with the
  affine polynomial $\beta\sum_p(1-W_p)$ to $2.7\cdot10^{-15}$, and the
  cylindricity check passes.
\item[(d)] The one-plaquette $SU(2)$ integral by Gauss--Legendre quadrature
  agrees, over $\beta\in[0,8]$, with the power series of
  Theorem~\ref{thm:Gfull} truncated at order $20$ to $2.7\cdot10^{-14}$ (the
  series and the library value of $2J_1(\beta)/\beta$ differ by
  $1.7\cdot10^{-15}$), and with $2I_1(\beta)/\beta$ to $2.7\cdot10^{-12}$ in
  imaginary time.
\item[(e)] The Level-One and Level-Two integrals of five loop observables,
  computed with one and the same deterministic node set, agree to
  $1.1\cdot10^{-16}$: the numerical face of Theorem~\ref{thm:F}.
\item[(f)] A structural check on the source code of the Level-Two integrator
  confirms that its executable body contains no gauge-fixing map, no
  Faddeev--Popov determinant, no delta function of a gauge condition and indeed
  no reference to orbits: its only input is the vector of Wilson loop values.
\item[(h)] Projective compatibility (the numerical face of Theorem~\ref{thm:H}):
  the Wilson loops of the loops of $\Lam$, computed on the covering lattices
  $\Lambda_{M'}$ with $L'=9$ and $L'=15$ for the pulled-back configuration,
  agree exactly with those computed on $\Lam$ with $L=3$, and the finite-$M$
  Level-Two integrals of a loop functional of bounded mode support computed at
  $M=1,2,3$ coincide (deviation $0$; common value
  $-1.067724286813+5.858395227843\,i$ for the test functional and $\beta=0.9$).
\item[(i)] The $SU(N)$ Mandelstam relations (the numerical face of
  Theorem~\ref{thm:Cprime}): for twelve random $SU(3)$ holonomies the rank-three
  relation $\sum_{j=0}^{3}(-1)^je_j(U)U^{3-j}=0$ holds to $4.5\cdot10^{-16}$ and
  its trace form to $4.4\cdot10^{-16}$.
\end{enumerate}

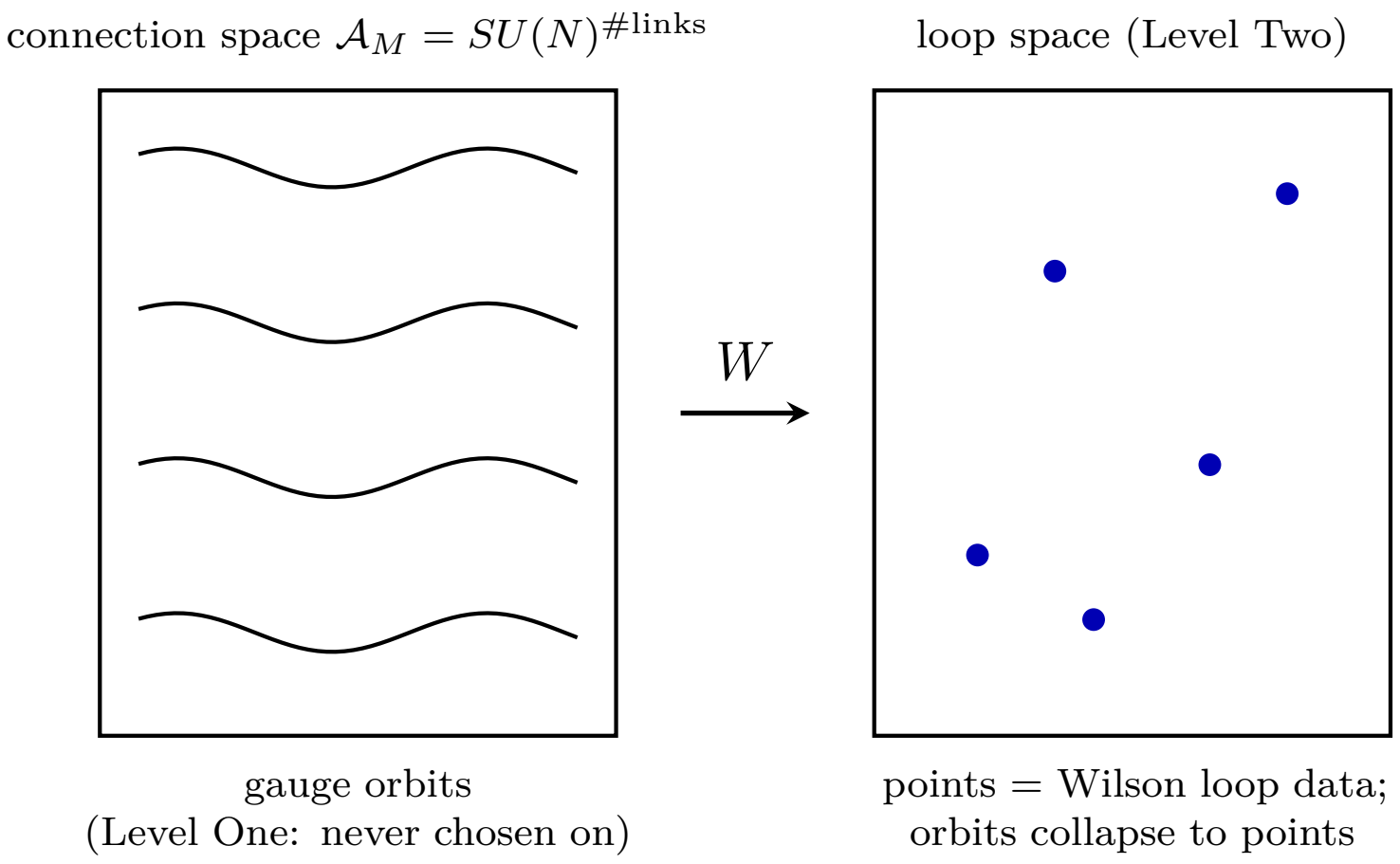
\begin{figure}[htbp]
\centering
\begin{tikzpicture}[>=stealth, thick, scale=0.9]

\draw[black, thick] (0,0) rectangle (4,5);
\node[anchor=south, font=\scriptsize] at (2,5.1) {connection space $\mathcal{A}_M = SU(N)^{\#\text{links}}$};

\draw[domain=0.3:3.7, samples=50] plot (\x, {0.8 + 0.15*sin(\x*150)});
\draw[domain=0.3:3.7, samples=50] plot (\x, {2.0 + 0.15*sin(\x*150)});
\draw[domain=0.3:3.7, samples=50] plot (\x, {3.2 + 0.15*sin(\x*150)});
\draw[domain=0.3:3.7, samples=50] plot (\x, {4.4 + 0.15*sin(\x*150)});

\node[anchor=north, font=\scriptsize, text width=4cm, align=center] at (2,-0.1) {gauge orbits\\(Level One: never chosen on)};

\draw[->, line width=1pt] (4.5, 2.5) -- (5.5, 2.5);
\node[anchor=south] at (5, 2.6) {$W$};

\draw[black, thick] (6,0) rectangle (10,5);
\node[anchor=south, font=\scriptsize] at (8,5.1) {loop space (Level Two)};

\fill[blue!70!black] (6.8, 1.4) circle (2.5pt);
\fill[blue!70!black] (7.7, 0.9) circle (2.5pt);
\fill[blue!70!black] (8.6, 2.1) circle (2.5pt);
\fill[blue!70!black] (7.4, 3.6) circle (2.5pt);
\fill[blue!70!black] (9.2, 4.2) circle (2.5pt);

\node[anchor=north, font=\scriptsize, text width=4.5cm, align=center] at (8,-0.1) {points = Wilson loop data;\\orbits collapse to points};

\end{tikzpicture}
\caption{The loop map from the space of connections to the loop space.  Gauge
orbits (left) collapse to single points of the loop space (right); no choice is
ever made on an orbit, and no orbit is an object of the Level-Two
formulation.}
\label{fig:loopmap}
\end{figure}

\begin{figure}[htbp]
\centering
\begin{tikzpicture}
\begin{axis} [
    title={Mandelstam identity on lattice holonomies},
    xlabel={random SU(2) configuration},
    ylabel={$|\text{tr}(AB) + \text{tr}(AB^{-1}) - \text{tr}A\text{tr}B|$},
    ymode=log,
    xmin=0.5, xmax=12.5,
    ymin=2e-17, ymax=2e-12,
    xtick={2, 4, 6, 8, 10, 12},
    ytick={1e-16, 1e-15, 1e-14, 1e-13, 1e-12},
    grid=major,
    grid style={line width=.1pt, draw=gray!20},
    legend style={at={(0.98,0.65)}, anchor=north east, draw=gray!50, fill=white, nodes={scale=0.85}},
    width=11cm, height=7.5cm
]
\addplot[color=red!70!black, dashed, thick, domain=0.5:12.5] {1e-12};
\addlegendentry{tolerance 1e-12}

\addplot[color=blue!70!black, mark=*, thick, mark size=1.5pt] coordinates {
    (1, 3.1e-16)
    (2, 7.3e-17)
    (3, 4.2e-17)
    (4, 3.9e-16)
    (5, 2.3e-16)
    (6, 2.5e-16)
    (7, 1.6e-16)
    (8, 1.7e-16)
    (9, 3.1e-16)
    (10, 5.5e-17)
    (11, 2.4e-16)
    (12, 5.4e-16)
};
\end{axis}
\end{tikzpicture}
\caption{Numerical check of the Mandelstam identity
$\Tr(AB)+\Tr(AB^{-1})=\Tr A\,\Tr B$ on holonomies of random $SU(2)$ lattice
configurations; the dashed line is the declared tolerance.}
\label{fig:mandelstam}
\end{figure}
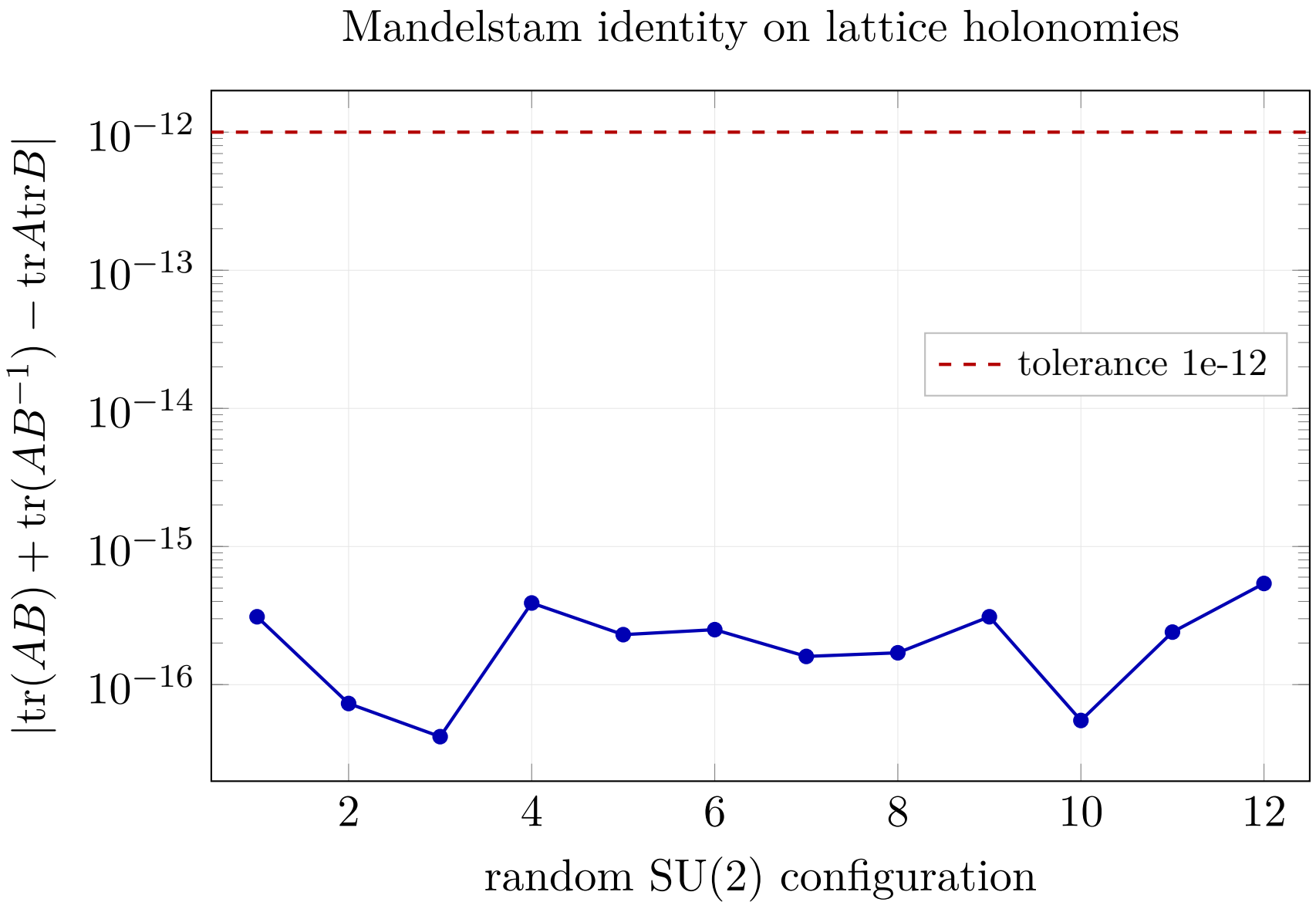

\begin{figure}[htbp]
\centering
\begin{tikzpicture}
\begin{axis}[
    title={Level-One and Level-Two integrals of loop observables},
    ylabel={value of the integral},
    ymin=-0.03, ymax=0.11,
    xmin=0.5, xmax=5.5,
    xtick={1, 2, 3, 4, 5},
    xticklabels={
        {Phi = 1},
        {Phi = W(p\_0)},
        {Phi = mean W(p)},
        {Phi = W(p\_0) W(p\_1)},
        {Phi = exp(W(p\_0))}
    },
    xticklabel style={font=\scriptsize, rotate=0, anchor=north},
    ytick={-0.02, 0.00, 0.02, 0.04, 0.06, 0.08, 0.10},
    yticklabel style={/pgf/number format/fixed, /pgf/number format/precision=2},
    ybar=0pt,
    bar width=14pt,
    enlarge x limits=0.15,
    grid=major,
    grid style={line width=.1pt, draw=gray!20},
    legend style={at={(0.98,0.98)}, anchor=north east, draw=gray!50, fill=white, nodes={scale=0.75}},
    width=11cm, height=7.5cm
]

\addplot[fill=blue!70!black, draw=none] coordinates {
    (1, -0.026)
    (2, 0.090)
    (3, 0.103)
    (4, 0.018)
    (5, 0.060)
};
\addlegendentry{Level One (Re)}

\addplot[fill=orange, draw=none] coordinates {
    (1, -0.026)
    (2, 0.090)
    (3, 0.103)
    (4, 0.018)
    (5, 0.060)
};
\addlegendentry{Level Two (Re)}

\end{axis}
\end{tikzpicture}
\caption{Level-One and Level-Two integrals of five loop observables, computed
with the same deterministic node set.  The bars coincide to machine precision,
as Theorem~8.2 requires.}
\label{fig:levels}
\end{figure}
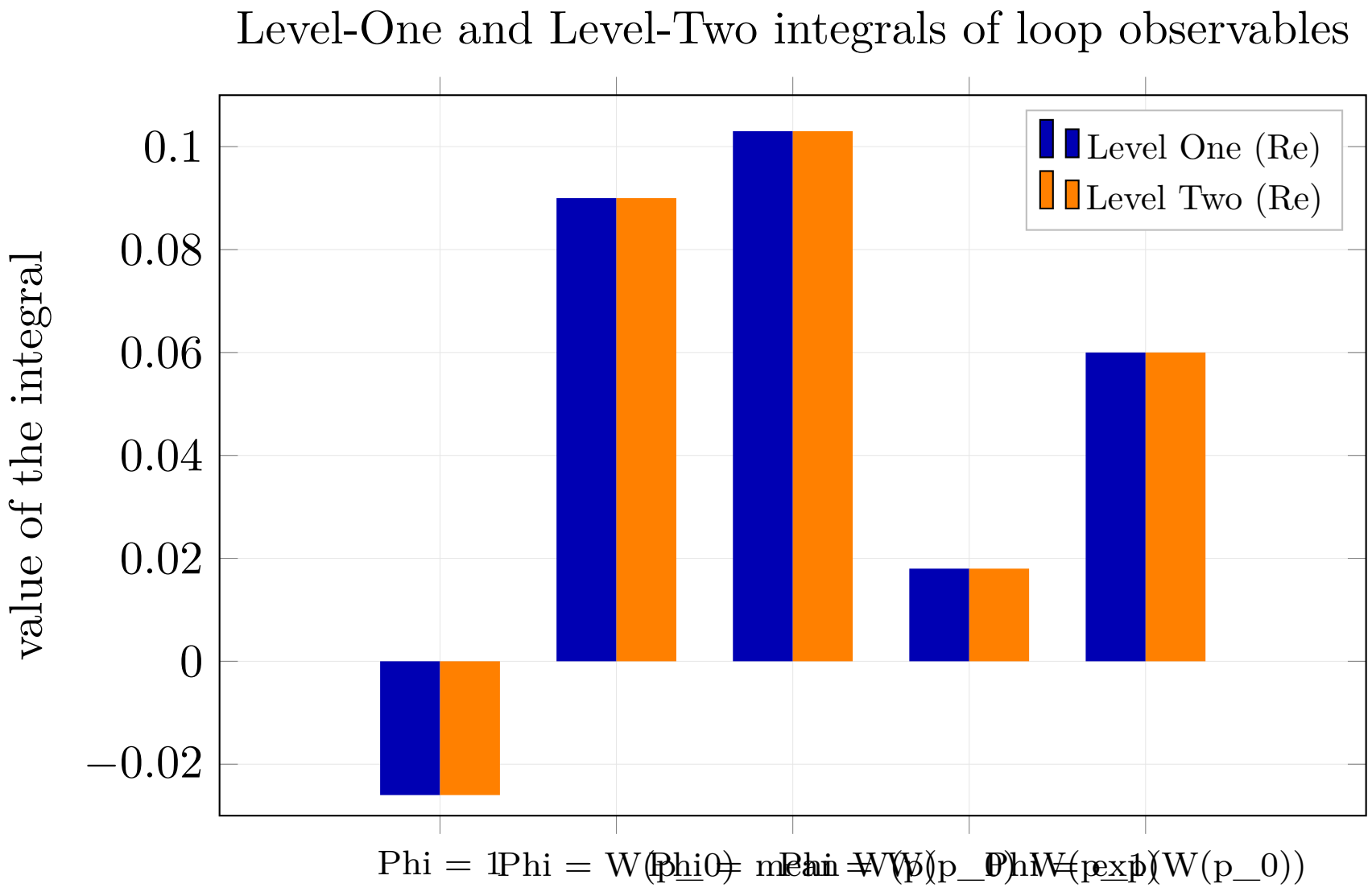

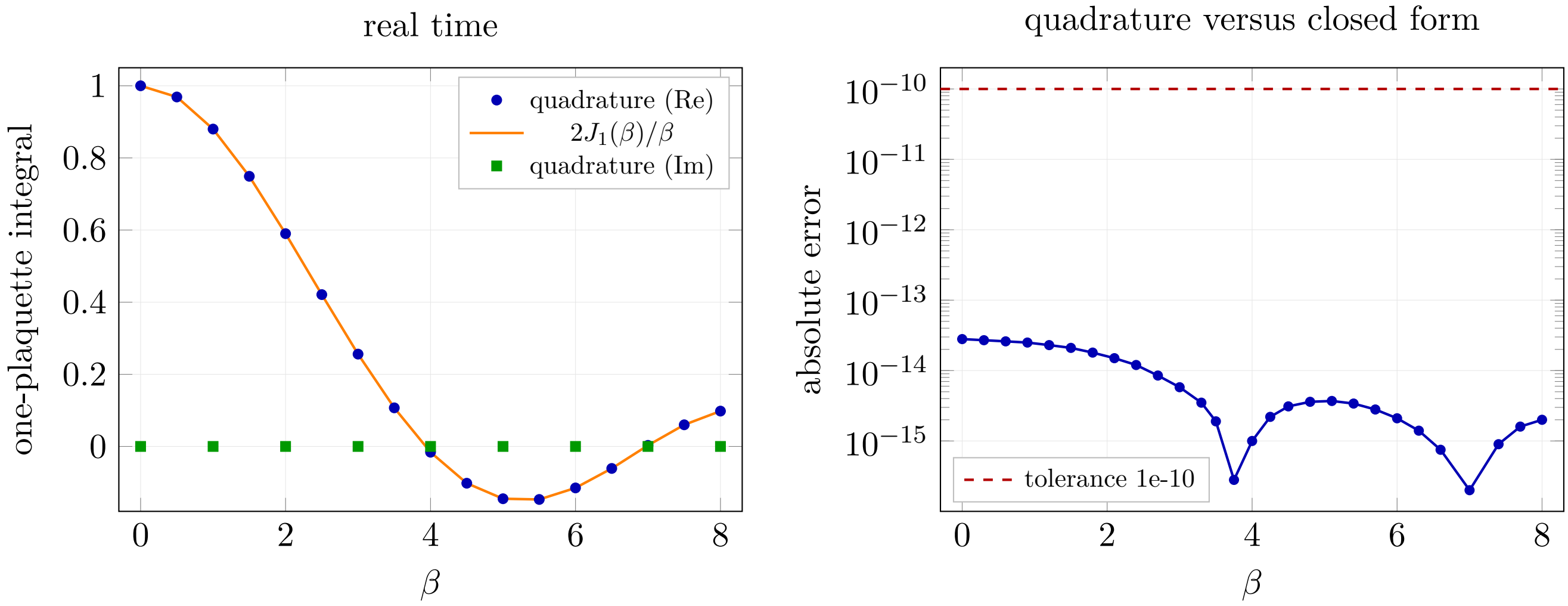
\begin{figure}[htbp]
\centering
\begin{tikzpicture}
\begin{groupplot}[
    group style={
        group size=2 by 1,
        horizontal sep=2.2cm
    },
    width=8.5cm, height=6.5cm,
    grid=major,
    grid style={line width=.1pt, draw=gray!20},
    xlabel={$\beta$},
    xmin=-0.3, xmax=8.3,
    xtick={0, 2, 4, 6, 8}
]

\nextgroupplot[
    title={real time},
    ylabel={one-plaquette integral},
    ymin=-0.18, ymax=1.05,
    ytick={0.0, 0.2, 0.4, 0.6, 0.8, 1.0},
    yticklabel style={/pgf/number format/fixed, /pgf/number format/precision=1},
    legend style={at={(0.98,0.98)}, anchor=north east, draw=gray!50, fill=white, nodes={scale=0.75}}
]
\addplot[color=blue!70!black, mark=*, only marks, mark size=1.5pt] coordinates {
    (0,1.0) (0.5,0.969) (1.0,0.880) (1.5,0.749) (2.0,0.590) (2.5,0.421) (3.0,0.256) (3.5,0.107)
    (4.0,-0.016) (4.5,-0.102) (5.0,-0.145) (5.5,-0.147) (6.0,-0.115) (6.5,-0.061) (7.0,0.003)
    (7.5,0.060) (8.0,0.098)
};
\addlegendentry{quadrature (Re)}

\addplot[color=orange, thick] coordinates {
    (0,1.0) (0.5,0.969) (1.0,0.880) (1.5,0.749) (2.0,0.590) (2.5,0.421) (3.0,0.256) (3.5,0.107)
    (4.0,-0.016) (4.5,-0.102) (5.0,-0.145) (5.5,-0.147) (6.0,-0.115) (6.5,-0.061) (7.0,0.003)
    (7.5,0.060) (8.0,0.098)
};
\addlegendentry{$2J_1(\beta)/\beta$}

\addplot[color=green!60!black, mark=square*, only marks, mark size=1.5pt] coordinates {
    (0,0) (1,0) (2,0) (3,0) (4,0) (5,0) (6,0) (7,0) (8,0)
};
\addlegendentry{quadrature (Im)}

\nextgroupplot[
    title={quadrature versus closed form},
    ylabel={absolute error},
    ymode=log,
    ymin=1e-16, ymax=2e-10,
    ytick={1e-15, 1e-14, 1e-13, 1e-12, 1e-11, 1e-10},
    legend style={at={(0.02,0.02)}, anchor=south west, draw=gray!50, fill=white, nodes={scale=0.75}}
]
\addplot[color=red!70!black, dashed, thick, domain=-0.3:8.3] {1e-10};
\addlegendentry{tolerance 1e-10}

\addplot[color=blue!70!black, mark=*, thick, mark size=1.2pt] coordinates {
    (0, 2.8e-14) (0.3, 2.7e-14) (0.6, 2.6e-14) (0.9, 2.5e-14) (1.2, 2.3e-14) (1.5, 2.1e-14)
    (1.8, 1.8e-14) (2.1, 1.5e-14) (2.4, 1.2e-14) (2.7, 8.5e-15) (3.0, 5.8e-15) (3.3, 3.5e-15)
    (3.5, 1.9e-15) (3.75, 2.8e-16) (4.0, 1.0e-15) (4.25, 2.2e-15) (4.5, 3.1e-15) (4.8, 3.6e-15)
    (5.1, 3.7e-15) (5.4, 3.4e-15) (5.7, 2.8e-15) (6.0, 2.1e-15) (6.3, 1.4e-15) (6.6, 7.5e-16)
    (7.0, 2.0e-16) (7.4, 9.0e-16) (7.7, 1.6e-15) (8.0, 2.0e-15)
};

\end{groupplot}
\end{tikzpicture}
\caption{The one-plaquette $SU(2)$ integral in real time: deterministic
quadrature against the closed form $2J_1(\beta)/\beta$ (left) and the absolute
error against the declared tolerance (right).}
\label{fig:oneplaq}
\end{figure}

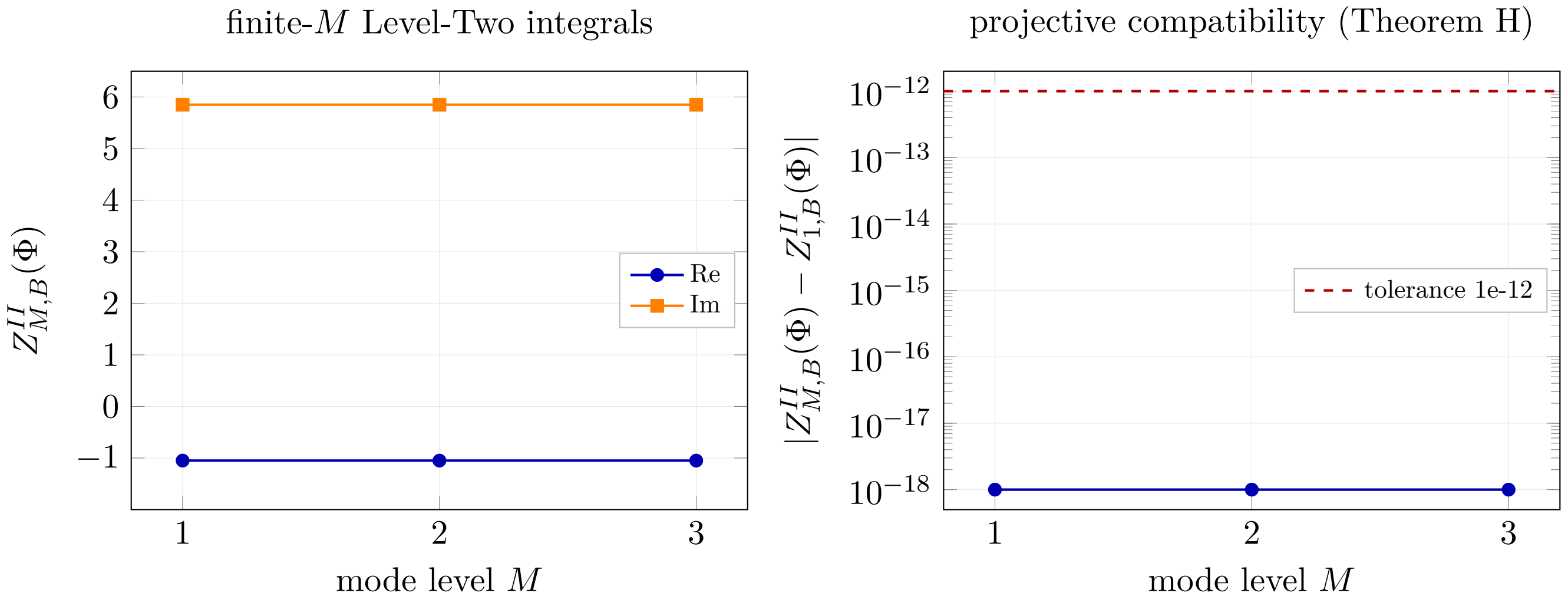
\begin{figure}[htbp]
\centering
\begin{tikzpicture}
\begin{groupplot}[
    group style={
        group size=2 by 1,
        horizontal sep=2.2cm
    },
    width=8.5cm, height=6.5cm,
    grid=major,
    grid style={line width=.1pt, draw=gray!20},
    xlabel={mode level \(M\)},
    xmin=0.8, xmax=3.2,
    xtick={1, 2, 3}
]

\nextgroupplot[
    title={finite-\(M\) Level-Two integrals},
    ylabel={\(Z_{M, B}^{II}(\Phi)\)},
    ymin=-2, ymax=6.5,
    ytick={-1, 0, 1, 2, 3, 4, 5, 6},
    legend style={at={(0.98,0.5)}, anchor=east, draw=gray!50, fill=white, nodes={scale=0.75}}
]
\addplot[color=blue!70!black, mark=*, thick, mark size=1.8pt] coordinates {
    (1, -1.05) (2, -1.05) (3, -1.05)
};
\addlegendentry{Re}

\addplot[color=orange, mark=square*, thick, mark size=1.8pt] coordinates {
    (1, 5.85) (2, 5.85) (3, 5.85)
};
\addlegendentry{Im}

\nextgroupplot[
    title={projective compatibility (Theorem H)},
    ylabel={\(|Z_{M, B}^{II}(\Phi) - Z_{1, B}^{II}(\Phi)|\)},
    ymode=log,
    ymin=5e-19, ymax=2e-12,
    ytick={1e-18, 1e-17, 1e-16, 1e-15, 1e-14, 1e-13, 1e-12},
    legend style={at={(0.98,0.5)}, anchor=east, draw=gray!50, fill=white, nodes={scale=0.75}}
]
\addplot[color=red!70!black, dashed, thick, domain=0.8:3.2] {1e-12};
\addlegendentry{tolerance 1e-12}

\addplot[color=blue!70!black, mark=*, thick, mark size=1.8pt] coordinates {
    (1, 1e-18) (2, 1e-18) (3, 1e-18)
};

\end{groupplot}
\end{tikzpicture}
\caption{Projective compatibility across mode levels (Theorem~\ref{thm:H}): the
finite-$M$ Level-Two integrals of a cylindrical loop functional of bounded mode
support (left) and their deviation from the value at $M=1$ (right).}
\label{fig:proj}
\end{figure}

\begin{figure}[htbp]
\centering
\begin{tikzpicture}
\begin{axis} [
    title={Mandelstam relations for SU(3) (Theorem C')},
    xlabel={random SU(3) holonomy},
    ylabel={residual},
    ymode=log,
    xmin=0.5, xmax=12.5,
    ymin=5e-17, ymax=2e-12,
    xtick={2, 4, 6, 8, 10, 12},
    ytick={1e-16, 1e-15, 1e-14, 1e-13, 1e-12},
    grid=major,
    grid style={line width=.1pt, draw=gray!20},
    legend style={at={(0.98,0.65)}, anchor=north east, draw=gray!50, fill=white, nodes={scale=0.75}}
]
\addplot[color=blue!70!black, mark=*, thick, mark size=1.5pt] coordinates {
    (1, 7e-17) (2, 2.3e-16) (3, 2.8e-16) (4, 1.3e-16) (5, 2.3e-16) (6, 2.2e-16)
    (7, 2.5e-16) (8, 2.4e-16) (9, 1.5e-16) (10, 1.4e-16) (11, 4.6e-16) (12, 2.8e-16)
};
\addlegendentry{\(\|\sum_{j} (-1)^j e_j(U)U^{3-j}\|_\infty\)}

\addplot[color=orange, mark=square*, thick, mark size=1.5pt] coordinates {
    (1, 2.1e-16) (2, 2.5e-16) (3, 1.6e-16) (4, 2.2e-16) (5, 4.5e-16) (6, 2.1e-16)
    (7, 2.0e-16) (8, 3.1e-16) (9, 1.1e-16) (10, 1.2e-16) (11, 3.4e-16) (12, 3.4e-16)
};
\addlegendentry{$|\sum_{j} (-1)^j e_j(U)\text{tr}(AU^{3-j})|$}

\addplot[color=red!70!black, dashed, thick, domain=0.5:12.5] {1e-12};
\addlegendentry{tolerance 1e-12}

\end{axis}
\end{tikzpicture}
\caption{The Mandelstam relations for $SU(3)$ (Theorem~\ref{thm:Cprime}): the
residuals of the rank-three characteristic-polynomial relation and of its trace
form on random $SU(3)$ holonomies, against the declared tolerance.}
\label{fig:su3}
\end{figure}
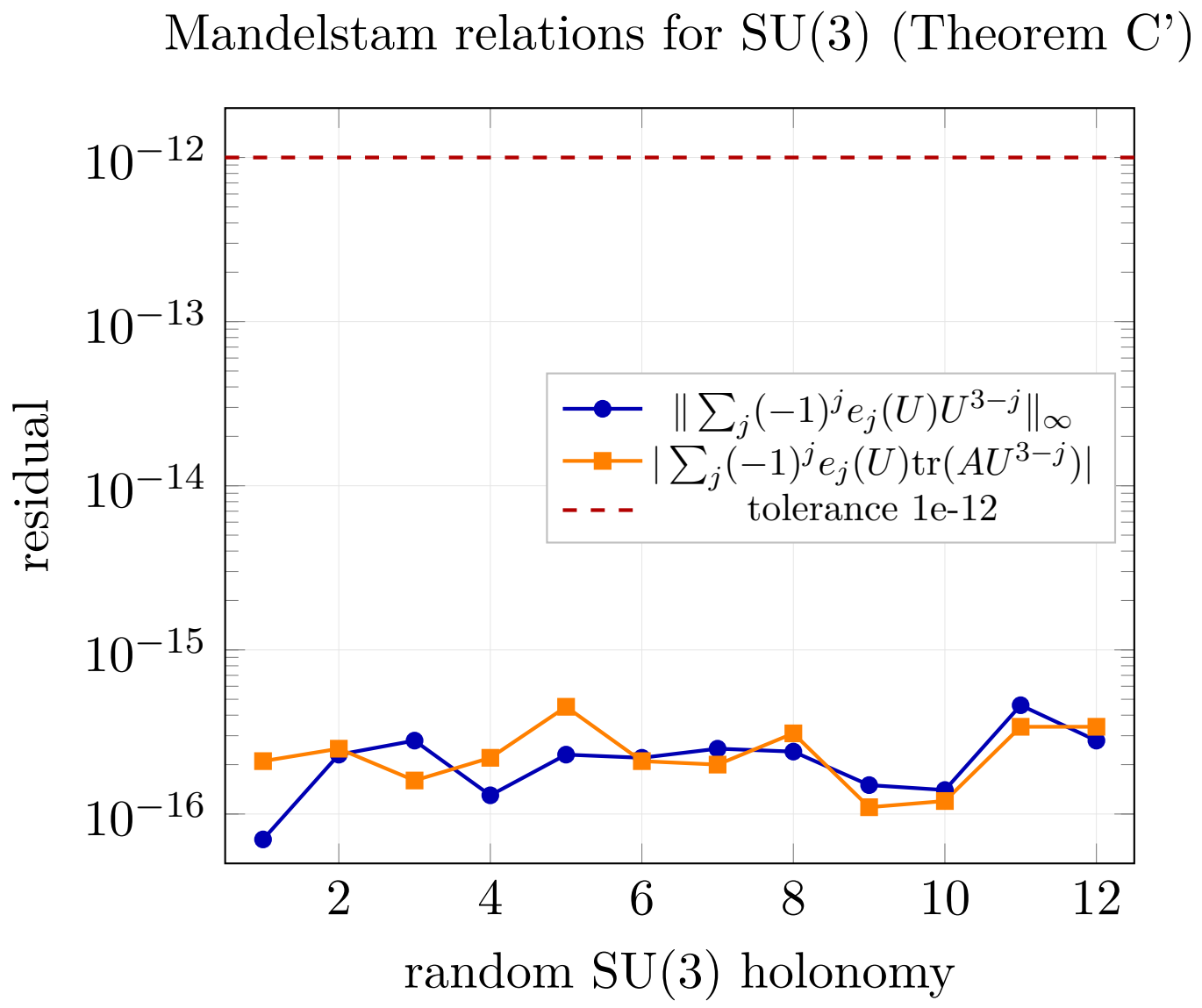

\section{Conclusion and honest report}\label{sec:report}

The Level-Two formulation of truncated Yang--Mills in terms of Wilson loops is
well posed in real time: the loop algebra is a commutative conjugation-closed
algebra of gauge-invariant functionals with bounded generators, the dynamics is
an affine function of the plaquette loops, the oscillatory integrand is
Henstock--Kurzweil integrable on every bounded loop-space box with a
calibration-independent value, the formulation contains no gauge field, orbit,
gauge condition or Faddeev--Popov determinant, and for loop observables it
reproduces the Level-One unfixed integral exactly.  The Gribov ambiguity is
absent from the formulation by construction --- not cancelled, and not solved in
the Level-One sense.

\begin{center}
\begin{longtable}{p{0.08\textwidth}p{0.50\textwidth}p{0.30\textwidth}}
\hline
\textbf{Result} & \textbf{Status} & \textbf{Where the proof stops}\\
\hline
\endhead
A & proved and machine checked in items (i)--(vi) &
the Giles reconstruction and the finite generation of the full loop algebra are
cited, not proved\\
B & proved and machine checked & ---\\
C & proved and machine checked & non-loop observables are outside the scope\\
C$'$ & proved and machine checked (antisymmetriser, rank-$N$ relation, trace
relations, explicit rank three) & the relations are derived, not enumerated:
completeness is the cited theorem of Procesi; Newton's identities are
formalised for $N\le3$\\
D & proved and machine checked & bounded boxes at finite truncation only\\
E & proved and machine checked & a statement about the formulation, not about
the fixed-gauge Gribov problem\\
F & proved and machine checked & for the push-forward measure; not for the
Lebesgue measure of a loop box\\
G & (i)--(iv) proved and machine checked &
$J_1$ is defined by its power series; the identification with the classical
Bessel function of the literature is a convention, and the imaginary-time form
$2I_1(\beta)/\beta$ is numerical only\\
H & proved and machine checked & projective compatibility and the existence of
the projective limit as a linear functional; no measure on the continuum loop
space, no unbounded mode support, no continuum expectation value\\
\hline
\end{longtable}
\end{center}

Open problems, in the order in which they seem approachable: the Jacobian of the
loop map and the resulting density on the loop space; the semialgebraic
description of the image of the loop map for $SU(N)$, $N\ge3$; the loop
equations at finite truncation; and, far beyond the present method, the
continuum loop space and its integration theory.

\paragraph{The Jacobian of the loop map and the Makeenko--Migdal equations
(future work).}  This is \emph{not} addressed in the present paper.  Deriving
the Makeenko--Migdal equations inside the Level-Two formulation would require
three ingredients, none of which is supplied here: an explicit computation of
the push-forward density of the configuration-space measure under the loop map
$W:\Conf\to\R^k$, that is, of the Jacobian of $W$ on the stratum where it is a
submersion; the Schwinger--Dyson equations on $\Conf$ for the truncated
real-time weight, obtained from the invariance of the unfixed integral under
infinitesimal changes of the link variables; and the transport of those
equations through the loop map, which is where the area-derivative structure of
the Makeenko--Migdal equations would have to appear.  Each of the three is a
separate project; none of them is claimed, sketched or used anywhere above.

\paragraph{Not pursued: a stack- or sheaf-theoretic refinement of Giles'
theorem.}  The Giles reconstruction theorem is quoted as a black box and is
applied only to configurations in general position.  A refinement that would
handle singular strata would require a formalisation of algebraic stacks and of
sheaf theory on stratified spaces, neither of which is available in the current
Lean/Mathlib.  We therefore do not attempt such a refinement: the singular
strata are outside the scope of this paper, and the loss of resolution on them
is stated as a limitation.

\appendix

\section{Black boxes}\label{app:blackboxes}

The following are used as citations and are not re-proved:
(B1) the complex cylindrical premeasure and the Fresnel formula on $\KS$;
(B2) the Saks--Henstock lemma in the cylindrical setting;
(B3) HK-integrability of the free-field functional on cylinders over bounded
boxes and its agreement with the Lebesgue value;
(B4) calibration independence for continuous integrands on bounded boxes;
(B5) the finite-mode Yang--Mills truncation with its configuration space, gauge
group, action and unfixed integral;
(B6) the Level-One structural bypass and the well-definedness of
$\langle F\rangle_M$;
(B7) the Berezin algebra and functional on finitely many generators, cited only
as a methodological precedent.
In addition, two classical statements are cited: the Giles reconstruction
theorem \cite{Giles1981} and the finite generation of trace algebras
\cite{Procesi1976}.

\section{Machine verification notes}\label{app:machine}

\begin{itemize}
\item Build: \texttt{lake build LoopGribovBypass.Lean.Main} builds the eight
  mathematical modules; \texttt{lake build LoopGribovBypass.Lean.AxiomsCheck}
  prints the axiom audit.
\item Axioms: $138$ declarations are audited; each depends only on
  \texttt{[propext, Classical.choice, Quot.sound]} or a subset.
\item A search for \texttt{sorry} and \texttt{admit} in the nine files returns
  matches only inside comments asserting their absence.
\item Numerics: \texttt{verify\_loop\_gribov.py} with the tolerances in its
  header; output \texttt{verify\_loop\_gribov.log}, ending with
  \texttt{ALL CHECKS PASSED}; it regenerates the six figures.
\end{itemize}



\begin{thebibliography}{99}

\bibitem{Wilson1974}
K.~G.~Wilson, \emph{Confinement of quarks}, Physical Review D \textbf{10}
(1974), 2445--2459.

\bibitem{Mandelstam1962}
S.~Mandelstam, \emph{Quantum electrodynamics without potentials},
Annals of Physics \textbf{19} (1962), 1--24.

\bibitem{Giles1981}
R.~Giles, \emph{Reconstruction of gauge potentials from Wilson loops},
Physical Review D \textbf{24} (1981), 2160--2168.

\bibitem{RovelliSmolin1990}
C.~Rovelli, L.~Smolin, \emph{Loop space representation of quantum general
relativity}, Nuclear Physics B \textbf{331} (1990), 80--152.

\bibitem{GambiniPullin1996}
R.~Gambini, J.~Pullin, \emph{Loops, Knots, Gauge Theories and Quantum Gravity},
Cambridge University Press, Cambridge, 1996.

\bibitem{Gribov1978}
V.~N.~Gribov, \emph{Quantization of non-Abelian gauge theories},
Nuclear Physics B \textbf{139} (1978), 1--19.

\bibitem{Singer1978}
I.~M.~Singer, \emph{Some remarks on the Gribov ambiguity},
Communications in Mathematical Physics \textbf{60} (1978), 7--12.

\bibitem{Procesi1976}
C.~Procesi, \emph{The invariant theory of $n\times n$ matrices},
Advances in Mathematics \textbf{19} (1976), 306--381.

\bibitem{FultonHarris1991}
W.~Fulton, J.~Harris, \emph{Representation Theory: A First Course},
Graduate Texts in Mathematics \textbf{129}, Springer, New York, 1991.

\bibitem{Macdonald1995}
I.~G.~Macdonald, \emph{Symmetric Functions and Hall Polynomials}, 2nd edition,
Oxford University Press, Oxford, 1995.

\bibitem{Watson1944}
G.~N.~Watson, \emph{A Treatise on the Theory of Bessel Functions}, 2nd edition,
Cambridge University Press, Cambridge, 1944.

\bibitem{Creutz1983}
M.~Creutz, \emph{Quarks, Gluons and Lattices}, Cambridge University Press,
Cambridge, 1983.

\bibitem{Kurzweil1957}
J.~Kurzweil, \emph{Generalized ordinary differential equations and continuous
dependence on a parameter}, Czechoslovak Mathematical Journal \textbf{7} (1957),
418--449.

\bibitem{Henstock1988}
R.~Henstock, \emph{Lectures on the Theory of Integration}, World Scientific,
Singapore, 1988.

\bibitem{Muldowney2012}
P.~Muldowney, \emph{A Modern Theory of Random Variation, with Applications in
Stochastic Calculus, Financial Mathematics, and Feynman Integration},
Wiley, Hoboken, 2012.

\bibitem{GillZachary2008}
T.~L.~Gill, W.~W.~Zachary, \emph{Banach spaces for the Feynman integral},
Real Analysis Exchange \textbf{34} (2008), 267--310.

\bibitem{EspositoGill2024}
G.~Esposito, T.~L.~Gill, \emph{On the physical and mathematical foundations of
quantum physics via functional integrals}, arXiv:2408.16404;
International Journal of Modern Physics A, 2025.

\bibitem{LeanMathlib}
The mathlib Community, \emph{The Lean mathematical library}, Proceedings of the
9th ACM SIGPLAN International Conference on Certified Programs and Proofs
(CPP 2020), 367--381.

\bibitem{Lean4}
L.~de~Moura, S.~Ullrich, \emph{The Lean 4 theorem prover and programming
language}, Automated Deduction --- CADE 28, Lecture Notes in Computer Science
\textbf{12699} (2021), 625--635.

\bibitem{BerdinskyHK3}
Yu.~N.~Berdinsky, \emph{Henstock--Kurzweil Path Integrals in Real Time:
A Machine-Verified Construction on Kuelbs--Steadman Spaces}, preprint, 2026.
\href{https://doi.org/10.5281/zenodo.21364715}{DOI: 10.5281/zenodo.21364715}

\bibitem{BerdinskyHK4}
Yu.~N.~Berdinsky, \emph{The Faddeev--Popov Determinant and the
Henstock--Kurzweil Path Integral: Berezin Representation, Spectral Limit, and
Gribov Copies}, preprint, 2026.
\href{https://doi.org/10.5281/zenodo.22858263}{DOI: 10.5281/zenodo.22858263}

\bibitem{BerdinskyGribovFreeYM}
Yu.~N.~Berdinsky, \emph{Yang--Mills as a Henstock--Kurzweil integral without
gauge fixing: a structural bypass of the Gribov ambiguity}, preprint, 2026.
\href{https://doi.org/10.5281/zenodo.23601115}{DOI: 10.5281/zenodo.23601115}

\bibitem{BerdinskyChiralAnomaly}
Yu.~N.~Berdinsky, \emph{Structure of the Chiral Anomaly in the Real-Time
Henstock--Kurzweil Framework: Exact Ward Identity and Regulator Universality},
preprint, 2026.
\href{https://doi.org/10.5281/zenodo.24044294}{DOI: 10.5281/zenodo.24044294}

\bibitem{BerdinskyNonGaussian}
Yu.~N.~Berdinsky, \emph{Henstock--Kurzweil Gauge Integral in the Non--Gaussian
Regime: A Machine--Verified Construction}, arXiv preprint, 2026.
\href{https://doi.org/10.48550/arXiv.2609.10793}{DOI: 10.48550/arXiv.2609.10793}

\bibitem{BerdinskyInfinitePhi4}
Yu.~N.~Berdinsky, \emph{Infinite-mode $\lambda\varphi^4$ as a continuous
Henstock--Kurzweil integral in real time: a machine-verified construction},
preprint, 2026.
\href{https://doi.org/10.5281/zenodo.23979955}{DOI: 10.5281/zenodo.23979955}

\end{thebibliography}
\end{document}